\documentclass[seceqn,dvips]{arxstspdf}
\usepackage{algorithm, dcolumn}
\usepackage{graphicx}
\usepackage{flushend}
\usepackage{stfloats}

\volume{23}
\issue{3}
\pubyear{2008}
\firstpage{383}
\lastpage{403}
\doi{10.1214/08-STS266}

\makeatletter

\newcolumntype{d}[1]{D{.}{.}{#1}}
\newtheorem{theorem}{Theorem}
\newtheorem{lemma}{Lemma}
\newtheorem{proposition}{Proposition}
\newproclaim{example}{Example}

\makeatother

\begin{document}
\begin{frontmatter}

\title{Multiway Spectral Clustering: A~Margin-Based Perspective}
\runtitle{Multiway Spectral Clustering}

\begin{aug}
\author[a]{\fnms{Zhihua} \snm{Zhang}\ead[label=e1]{zzhang@stat.berkeley.edu}}
\and
\author[a]{\fnms{Michael I.} \snm{Jordan}\corref{}\ead[label=e2]{jordan@stat.berkeley.edu}}
\runauthor{Z. Zhang and M. I. Jordan}

\affiliation{University of California, Berkeley}

\address[a]{Department of Statistics and Department of
Electrical Engineering and Computer Science, University of~California,
Berkeley, USA,
\printead{e1,e2}.
}

\end{aug}

% ABSTRACT
%
\begin{abstract}
Spectral clustering is a broad class of clustering procedures
in which an intractable combinatorial optimization formulation of
clustering is ``relaxed'' into a tractable eigenvector problem,
and in which the relaxed solution is subsequently ``rounded''
into an approximate discrete solution to the original problem.
In this paper we present a novel margin-based perspective on multiway
spectral clustering. We show that the margin-based perspective
illuminates both the relaxation and rounding aspects of spectral
clustering, providing a unified analysis of existing algorithms
and guiding the design of new algorithms. We also present
connections between spectral clustering and several other topics
in statistics, specifically minimum-variance clustering, Procrustes
analysis and Gaussian intrinsic autoregression.
\end{abstract}

% KEYWORDS
%

\end{frontmatter}

%s1 ###
\section{Introduction}

Spectral clustering is a promising approach to clustering that has
recently been undergoing rapid development (\cite*{Shi2000}; \cite*{Kannan};
\cite*{ZhaNIPS2002}; \cite*{Ng2002}; \cite*{MeilaUAI2005}; \cite*{DingSDM2005};
\cite*{Bach2006}; \cite*{LuxburgSC2007}).
In the spectral framework a clustering problem is posed as a discrete
optimization problem (an integer program). This problem is generally
intractable computationally, and approximate solutions are obtained
by a two-step procedure in which (1) the problem is ``relaxed'' into
a simplified continuous optimization problem that can be solved
efficiently, and (2) the resulting continuous solution is ``rounded''
into an approximate solution to the original discrete problem. The
adjective ``spectral'' refers to the fact that the relaxed problem
generally takes the form of an eigenvector problem (the original
objective function involves quadratic constraints, which yields a
Rayleigh coefficient in the relaxed problem).

The solutions of the relaxed problem are often referred to as
\textit{spectral embeddings} and have applications outside of the
clustering context (\cite*{Belkin2002}). Our focus here,
however, will be on spectral clustering.

Spectral clustering was first developed in the context of graph
partitioning problems (\cite*{Donath1973}; \cite*{Fiedler1973}),
where the problem is to partition a weighted graph into disjoint pieces,
minimizing the sum of the weights of the edges linking the disjoint pieces.
The methodology is applied to data analysis problems by identifying
nodes of the graph with data points and identifying the edge weights
with the similarity (or ``distance'') function used in clustering.
The problem then is to choose an appropriate relaxation of the
weighted graph partitioning problem and an appropriate rounding
procedure. The current literature offers many such
choices (see, e.g., \cite*{LuxburgSC2007}).

Naive formulations of graph cut problems yield uninteresting
solutions in which single nodes are separated from the rest of the
graph. The spectral \mbox{formulation} becomes interesting (and
computationally intractable) when some sort of constraint is imposed
so that the partition is balanced. There have been two main
approaches to imposing balancing constraints. In the \textit{ratio
cut} (\textsc{Rcut}) formulation (\cite*{ChanCUT1994}), the constraints are
expressed in terms of cardinalities of subsets of nodes. In the
\textit{normalized cut} (\textsc{Ncut}) formulation (\cite*{Shi2000}),
the constraints are expressed in terms of the degrees of
nodes. In this paper we study a general \textit{penalized cut} (\textsc{Pcut})
formulation that includes \textsc{Rcut} and \textsc{Ncut} as special cases
and we emphasize the close relationships between the spectral
relaxations resulting from \textsc{Rcut} and \textsc{Ncut} formulations.

A seemingly very different approach to clustering is the classical
minimum-variance formulation where one minimizes the trace of the
pooled within-class covariance matrix (\cite*{Webb2002}). As we
show, however, this formulation is closely related to \textsc{Pcut}.
In particular, posing the minimum-variance problem in the
reproducing kernel Hilbert space (RKHS) defined by a kernel
function (\cite*{Wahba1990}), we establish a connection between
spectral relaxation and minimum-variance clustering by treating the
Laplacian matrix in the \textsc{Pcut} formulation as the Moore--Penrose
inverse of the kernel matrix in the minimum-variance formulation.

Other forms of clustering procedures have been usefully analyzed
in terms of their relationships to discrimination or classification
procedures\break (\cite*{Webb2002}), and in the current paper we aim to
develop connections of this kind in the case of spectral clustering.
In this regard, it is important to note that our focus is on the
\textit{multiway} clustering problem, in which a data set is directly
partitioned into $c$ sets where $c > 2$. This differs from the
classical graph-partitioning literature, where the focus has been
on algorithms that partition a graph into two pieces (``binary cuts''),
with the problem of partitioning a graph into multiple pieces
(``multiway cuts'') often approached by the recursive invocation
of a binary cut algorithm.

In the case of binary cuts, an interesting connection to classification
has been established by \citet{RahimiRecht2004}, who have noted
that \textsc{Ncut}-based spectral clustering can be interpreted as
finding a hyperplane in an RKHS that falls in a ``gap'' in the
empirical distribution. In the current paper we show that this
idea can be extended to general multiway \textsc{Pcut} spectral
relaxation, where the intuitive idea of a ``gap'' can be expressed
precisely using ideas from the classification literature, specifically
the idea of a multiclass margin.

Turning to the rounding problem, we note first that for binary
cuts the rounding problem is a relatively simple problem, generally
involving the choice of a threshold for the elements of an
eigenvector (\cite*{Juhasz1977}; \cite*{Weiss2001}). The problem
is significantly more complex in the multiway case, however,
where it essentially involves an auxiliary \mbox{clustering} problem
based on the spectral embedding. For example, \citet{YuICCV2003} proposed
a rounding scheme that works with an alternative iteration between
singular value decomposition (SVD) and nonmaximum suppression,
whereas \citet{Bach2006} devised $K$-means and\break weighted $K$-means
algorithms for rounding. In the current paper we show that rounding
can be usefully approached within the framework of Procrustes
analysis (\cite*{GowerBook2004}). Moreover, we show that this
approach again reveals links between spectral methods and multiway
classification; in particular, we show that the auxiliary Procrustes
problem that we must solve can be analyzed using the tools of
margin-based classification.

Extant multiway spectral algorithms, including\break those of
\citet{Bach2006} and \citet{YuICCV2003}, as well as many
others (\cite*{Ng2002}; \cite*{ZhaNIPS2002}; \cite*{DingSDM2005}; \cite*{MeilaUAI2005}),
are based on the representation of spectral embeddings as
$c$-dimensional vectors. The redundancy inherent in using
$c$-dimensional vectors is inconvenient, however, preventing the
flow of results from the binary case to the multiway
case (\cite*{Shi2000}). The margin-based perspective that we pursue
here shows the value of working with a nonredundant,
$(c - 1)$-dimensional representation of the spectral embedding.

Our overall approach to spectral clustering is as follows. We first
construct a nonredundant, margin-based representation of multiway
spectral relaxation problems. Such a margin-based spectral relaxation
is a tractable constrained eigenvalue problem. We then carry out
a rounding scheme by solving an auxiliary Procrustes problem,
which is again associated with a margin-based classification
method. We refer to the resulting clustering framework---margin-based
spectral relaxation with margin-based rounding---as \textit{margin-based
spectral clustering}.

The margin-based approach not only provides substantial insight into
the relationships among spectral clustering procedures, but it also
yields probabilistic interpretations of these procedures.
Specifically, we show that the spectral relaxation obtained from the
\textsc{Pcut} framework can be interpreted as a form of Gaussian intrinsic
autoregression (\cite*{BesagBio1995}). These are limiting forms of
Gaussian conditional
autoregressions (\cite*{Besag1974}; \cite*{MardiaJMA1988}) that retain the
Markov property (two vertices in a graph are not connected if and
only if their corresponding embeddings in the intrinsic
autoregression are conditionally independent).

In summary, the current paper develops a mathematical perspective
on spectral clustering that unifies the various algorithms that have
been studied and emphasizes connections to other areas of statistics.
\mbox{Specifically} we discuss connections to multiway classification,
reproducing kernel Hilbert space methods, Procrustes analysis and
Gaussian intrinsic autoregression.

The remainder of the paper is organized as follows.
Sections~\ref{secpcutrelax} and~\ref{secvarrelax} describe multiway
spectral relaxation problems based on the general \textsc{Pcut} formulation
and the minimum variance formulation, respectively. The relationship
between these two formulations is also discussed in Section~\ref{secvarrelax}.
In Section~\ref{secrounding} we present two rounding schemes, one based
on Procrustean transformation and the other based on $K$-means.
We present a geometric perspective on spectral clustering using margin-based
principles in Section~\ref{secmargin}, and we discuss the connection
to Gaussian intrinsic autoregression models in Section~\ref{secgias}.
Experimental comparisons are given in Section~\ref{secexp2} and we
present our conclusions in Section~\ref{secconclusion}. Note that
several proofs are deferred to the \hyperref[ap:relaxation]{Appendix}.

We use the following notation in this paper.
$\mathbf{I}_{m}$ denotes the
$m\times m$ identity matrix,
$\mathbf{1}_m$ the
$m\times 1$ of ones, $\mathbf{0}$
the zero vector or matrix zero of appropriate size and
$\mathbf{H}_m = \mathbf{I}_m-\frac{1}{m} \mathbf{1}_m \mathbf{1}_m'$ the $m\times m$
centering matrix. For an $n\times 1$ vector
$\mathbf{a}=(a_1, \ldots, a_n)'$,
$\operatorname{diag}(\mathbf{a})$
represents the $n\times n$ diagonal matrix with
$a_{1}, \ldots, a_{n}$
as its diagonal entries and $\|\mathbf{a}\|$ is the Euclidean norm of
$\mathbf{a}$. For an
$m\times m$ matrix $\mathbf{A}=[a_{ij}]$, we let
$\operatorname{dg}(\mathbf{A})$ be
the diagonal matrix with $a_{11}, \ldots, a_{mm}$ as its diagonal
entries, $\mathbf{A}^{+}$ be the Moore--Penrose inverse of~$\mathbf{A}$, $\operatorname{tr}(\mathbf{A})$ be
the trace of~$\mathbf{A}$, $\operatorname{rk}(\mathbf{A})$ be the
rank of $\mathbf{A}$ and $\|\mathbf{A}\|_F$ be
the Frobenius norm of $\mathbf{A}$.

%%%%%%%%%%%%%%%%%%%%%%%%%%%%%%%%%%%%%%%%%%%%%%%%%%%%%%%%%%%%%%%%%%%%%%%%%%%%%%%%%%%%%%%%
%%%%%%%%%%%%%%%%%%%%%%%%%%%%%%%%%%%%%%%%%%%%%%%%%%%%%%%%%%%%%%%%%%%%%%%%%%%%%%%%%%%%%%%%
%s2 ###
\section{Spectral Relaxation for Penalized~Cuts} \label{secpcutrelax}

Given a set of $n$ $d$-dimensional data points,
$\{\mathbf{x}_1, \ldots,\break\mathbf{x}_n\}$, our goal is to cluster
the $\mathbf{x}_{i}$ into $c$ disjoint classes such that each
$\mathbf{x}_i$ belongs to one and only one class. We consider a graphical
representation of this problem. Let $V=\{1, 2, \ldots, n\}$ denote
the index set of the data points and consider an undirected
graph $\mathcal{G} = (V, \mathcal{E})$ where $V$ is the set
of nodes in the graph and $\mathcal{E}$ is the set of edges. Associated
with the graph is a symmetric $n\times n$ \textit{affinity matrix}
(also referred to as a \textit{similarity matrix}), $\mathbf{W}=[w_{ij}]$,
defined on pairs of indices such that $w_{ij} \geq0$ for $(i,j)
\in\mathcal{E}$ and $w_{ij} = 0$ otherwise. The values $w_{ij}$
are often obtained via a function evaluated on the corresponding pairs of
data vectors; that is, $w_{ij} = \psi(\mathbf{x}_i, \mathbf{x}_j)$ for some (symmetric)
function $\psi$. A~variety of different ways to map a data set into
a graph $\mathcal{G}$ and an affinity matrix $\mathbf{W}$ have been
explored in
the literature; for a review see von Luxburg~(\citeyear{LuxburgSC2007}).

The problem is thus to partition $V$ into $c$ subsets
$V_j$; that is, $V_i \cap V_j =\varnothing$ for $i\neq j$ and
$\bigcup_{j=1}^c V_j = V$, where the cardinality of $V_j$ is
$n_j$ so that $\sum_{j=1}^c n_j =n$. This problem is typically
formulated as a\break combinatorial optimization problem.
Let $W(A, B) = \sum_{i \in A, j \in B} w_{ij}$ for two
(possibly overlapping) subsets $A$ and $B$ of $V$ and
consider the following \textit{multiway penalized cut} criterion:
% XXX check that the \sc comes out OK in the equation

%
%e2.1 ###
\begin{eqnarray} \label{eqpcut}
\mbox{\textsc{Pcut}} = \sum_{j=1}^c \frac{W(V_j, V)-W(V_j, V_j)}{\sum_{i\in V_j}\pi_i},
\end{eqnarray}
where $\bolds{\pi}=(\pi_1, \ldots, \pi_n)'$
is a user-defined vector
of weights (examples are provided below) with $\pi_i > 0$ for all $i$.
The numerator of each of the terms in this expression is equal
to the sum of the affinities on edges leaving the subset $V_j$.
Thus the minimization of \textsc{Pcut} with respect to the
partition $\{V_1, \ldots, V_c\}$ aims at finding a partition
in which edges with large affinities tend to stay within the
individual subsets $V_j$. The denominator weights $\sum_{i\in V_j}\pi_i$ encode
a notion of ``size'' of the subsets $V_j$ and act to balance the
partition.

The \textsc{Pcut} criterion can also be written in matrix notation
as follows. Define $\mathbf{D}=\operatorname{diag}(\mathbf{W}\mathbf
{1}_n)$ and let $\mathbf{L}= \mathbf{D}-\mathbf{W}$
denote the \textit{Laplacian matrix} of the graph. (An $n\times n$
matrix $\mathbf{L}=[l_{ij}]$ is a \textit{Laplacian matrix} if
$l_{ii}>0$ for
$i=1, \ldots, n$; $l_{ij} = l_{ji} \leq0$ for $i \neq j$;
$\sum_{j=1}^n l_{ij} =0$ for $i=1, \ldots, n$. Note that
Laplacian matrices are positive semidefinite (\cite*{Mohar91}).)
Let $\bolds{\Pi}=\operatorname{diag}(\pi_1, \ldots, \pi_n)$ be a
diagonal matrix
of weights. Let $t_i \in\{1, \ldots, c\}$ denote the
assignment of $\mathbf{x}_i$ to a cell in the partition and define
% XXX should there be a prime here or not?
the indicator matrix $\mathbf{E}=[\mathbf{e}_{1}, \ldots, \mathbf{e}_{n}]'$, where
$\mathbf{e}_{i} \in\{0, 1\}^{c\times 1}$ is a binary vector whose
$t_i$th entry is one and all other entries are zero.
It can now be readily verified that \textsc{Pcut} takes
the following form:
%
%e2.2 ###
\begin{equation}\label{eqpcutalt}
\mbox{\textsc{Pcut}} = \operatorname{tr}\bigl(\mathbf{E}'\mathbf{LE}(\mathbf{E}'\bolds{\Pi}\mathbf{E})^{-1}\bigr),
\end{equation}
where it is helpful to note that $(\mathbf{E}'\bolds{\Pi}\mathbf{E})^{-1}$ is a diagonal
matrix, implying that \textsc{Pcut} is simply a scaled quadratic
form. We wish to optimize this scaled \mbox{qua-}
dratic form with respect to $\mathbf{E}$.

Two well-known examples of the \textsc{Pcut} problem are the
\textit{ratio cut} (\textsc{Rcut}) problem~(\cite*{ChanCUT1994}),
in which $\bolds{\Pi}=\mathbf{I}_n$, and the \textit{normalized cut}
(\textsc{Ncut}) problem~(\cite*{Shi2000}), in which
$\bolds{\Pi}= \mathbf{D}$. In the \textsc{Rcut} problem the notion
of ``size'' of a subset $V_j$ is simply the number of nodes
in the subset, whereas in the \textsc{Ncut} problem ``size''
is captured by the total degree of the nodes in the subset.

The spectral clustering approach to minimizing \textsc{Pcut}
involves two stages: (1) we \textit{relax} the problem into a
tractable spectral analysis problem in which continuous variables
replace the indicators $\mathbf{E}$, and (2) we then employ a \textit{rounding}
scheme to obtain a partition $\{V_1, \ldots, V_n\}$ from the
continuous relaxation. In the remainder of this section, we
focus on the first step (the relaxation) and we return to the
rounding problem in Section~\ref{secrounding}.

The standard presentation of spectral relaxation proceeds
somewhat differently in the case of a binary partition and
a multiway partition (\cite*{LuxburgSC2007}). In both cases,
spectral relaxation is motivated by the observation that the
\textsc{Pcut} criterion in~(\ref{eqpcutalt}) has the form
of a Rayleigh coefficient, and that replacing the indicator
matrix $\mathbf{E}$ with a real-valued matrix yields a classical
generalized eigenvector problem. In the binary case,
the indicator matrix $\mathbf{E}$ has two columns, which yields
two generalized eigenvectors in the relaxed problem.
However, in the subsequent rounding procedure, the problem
is to discriminate between two classes, for which a single
vector direction suffices. To deal with this redundancy it
is standard to place a (linear) constraint upon the relaxation,
such that it is the second generalized eigenvector that is used
for rounding~(\cite*{LuxburgSC2007}). In the multiway case,
on the other hand, no such constraint is imposed; the
redundancy inherent in having $c$ generalized eigenvectors
to discriminate among $c$ classes is generally not addressed.
(It is resolved implicitly at the rounding stage.)

We find this distinction between the binary case and the
multiway case to be inconvenient, and thus in the approach
to be described in the following section we adopt an idea
from the literature on multiway
classification (e.g., \cite*{ZouZhuHastieTR2006}; \cite*{ShenWang07})
where nonredundant, $(c - 1)$-dimensional vectors are used to
discriminate among $c$ classes. These vectors are referred
to as \textit{margin vectors}. We refer the reader to the
classification literature for the geometric rationale behind
the terminology of ``margin'' (although we note that a geometric
interpretation of margin vectors will also appear in the current
paper in Section~\ref{sechyperplane}).

%s2.1 ###
\subsection{Spectral Relaxation}

To formulate a spectral relaxation of~(\ref{eqpcutalt}),
we replace the indicator matrix $\mathbf{E}$ with a real
$n\times (c - 1)$ matrix
$\mathbf{Y}=[\mathbf{y}_1, \ldots, \mathbf{y}_n]'$. The following
proposition,
which is based on a result of \citet{Bach2006}, shows that we
can express the \textsc{Pcut} criterion in terms of real-valued
matrices $\mathbf{Y}$ satisfying certain conditions.

\begin{proposition} \label{PRO1}
Let $\mathbf{Y}$ be an $n\times (c - 1)$ real matrix such that:
\textup{(a)} the columns of
$\mathbf{Y}$ are piecewise constant with respect to the partition
$\mathbf{E}$, \textup{(b)} $\mathbf{Y}' \bolds{\Pi}\mathbf{Y}=
\mathbf{I}_{c - 1}$
and \textup{(c)} $\mathbf{Y}' \bolds{\Pi}\mathbf{1}_n = \mathbf
{0}$. Then \textsc{Pcut} is equal to
$\operatorname{tr}\bigl(\mathbf{Y}' \mathbf{L}\mathbf{Y}\bigr)$.
\end{proposition}
The proof of Proposition~\ref{PRO1} is given in Appendix~\ref{aprelaxation}.

For this proposition to be useful it is necessary to show that
matrices satisfying the three conditions in Proposition~\ref{PRO1}
exist. Condition (a) for $\mathbf{Y}$ is equivalent to the statement
that $\mathbf{Y}$ can be expressed as $\mathbf{Y}=\mathbf{E}\bolds
{\Psi}$ where $\bolds{\Psi}$ is some
$c\times (c - 1)$ matrix. Thus, the question becomes whether
there exists a $\bolds{\Psi}$ such that $\mathbf{Y}$ satisfies
conditions~(b)--(c).
In Appendix~\ref{apexistence} we provide a general procedure
for constructing such a $\bolds{\Psi}$. This establishes the following
proposition.

\begin{proposition} \label{PROEXIST}
Matrices $\mathbf{Y}$ satisfying the three conditions in
Proposition~\ref{PRO1} exist.
\end{proposition}

We now obtain a spectral relaxation by dropping
condition (a). This yields the following optimization
problem:
\begin{eqnarray}\label{eqrelax1}
\nonumber&&\min_{\mathbf{Y}\in\mathbb{R}^{n\times(c - 1)}}\operatorname{tr}\bigl(\mathbf{Y}'
\mathbf{L}\mathbf{Y}\bigr) \\[-8pt]\\[-8pt]
\nonumber&&\quad\mbox{s.t. }\mathbf{Y}' \bolds{\Pi}\mathbf{Y}= \mathbf{I}_{c - 1}
\mbox{ and }{\mathbf{Y}}' \bolds{\Pi}\mathbf{1}_n = \mathbf{0},
\end{eqnarray}
which is a constrained generalized eigenvalue problem.

%%%%%%%%%%%%%%%%%%%%%%%%%%%%%%%%%%%%%%%%%%%%%%%%%%%%%%%%%%%%%%%%%%%%%
%%%%%%%%%%%%%%%%%%%%%%%%%%%%%%%%%%%%%%%%%%%%%%%%%%%%%%%%%%%%%%%%%%%%
%s2.2 ###
\subsection{Solving the Spectral Relaxation} \label{secrelax1}

Letting ${\mathbf{Y}}_0= \bolds{\Pi}^{1/2} \mathbf{Y}$, we
can transform (\ref{eqrelax1})
into the following problem:
\begin{eqnarray}\label{eqrelax2}
\nonumber&&\min_{\mathbf{Y}_0\in\mathbb{R}^{n\times (c - 1)}}
\operatorname{tr}(\mathbf{Y}_0'\bolds{\Pi}^{-1/2}
\mathbf{L}\bolds{\Pi}^{-1/2} {\mathbf{Y}}_0),
\\[-8pt]\\[-8pt]
\nonumber&&\quad\mbox{s.t. }{\mathbf{Y}}_0' {\mathbf{Y}}_0 = \mathbf{I}_{c - 1}
\mbox{ and }
{\mathbf{Y}}_0' \bolds{\Pi}^{1/2} \mathbf{1}_n = \mathbf{0}.
\end{eqnarray}
The solution to this constrained eigenvalue problem is given in
the following theorem.

\begin{theorem} \label{THMCEP0}
Suppose that $\mathbf{L}$ is a real symmetric matrix such that
$\mathbf{L}\mathbf{1}_n =\mathbf{0}$
and suppose that the diagonal entries of $\bolds{\Pi}$ are all
positive. Let
$\bolds{\mu}_1=\alpha\bolds{\Pi}^{1/2}\mathbf{1}_n$ be
the eigenvector associated
with the eigenvalue $\gamma_1=0$ of $\bolds{\Pi}^{-1/2}\mathbf{L}\bolds{\Pi}^{-1/2}$,
where $\alpha^2=1/(\mathbf{1}_n' \bolds{\Pi}\mathbf{1}_n)$.
Let the remaining eigenvalues of $\bolds{\Pi}^{-1/2}\mathbf{L}\bolds{\Pi}^{-1/2}$
be arranged so that $\gamma_2\leq\cdots\leq\gamma_n$, and let the
corresponding orthonormal eigenvectors be denoted by $\bolds{\mu}_i$,
$i=2, \ldots, n$. Then the solution of problem (\ref{eqrelax2}) is
$\bar{\mathbf{Y}}_0 = \mathbf{U}\mathbf{Q}$ where $\mathbf
{U}=[\bolds{\mu}_2, \ldots, \bolds{\mu}_c]$ and $\mathbf{Q}$ is
an arbitrary $(c - 1)\times (c - 1)$ orthonormal matrix, with
$\min\{\operatorname{tr}({\mathbf{Y}}_0' \bolds{\Pi}^{-1/2}\* \mathbf{L}\bolds{\Pi}^{-1/2}
{\mathbf{Y}}_0)\} = \sum_{i=2}^{c} \gamma_i$. Furthermore, if
$\gamma_c<\gamma_{c+1}$, then $\bar{\mathbf{Y}}_0$ is a strict local minimum of
$\operatorname{tr}({\mathbf{Y}}_0' \bolds{\Pi}^{-1/2}\*
\mathbf{L}\bolds{\Pi}^{-1/2} {\mathbf{Y}}_0)$.
\end{theorem}

It follows from the theorem that the solution of
problem~(\ref{eqrelax1}) is $\mathbf{Y}= \bolds{\Pi}^{-1/2} \mathbf{U}\mathbf{Q}$.
The proof of Theorem~\ref{THMCEP0} is given in Appendix~\ref{aplagrangian}.
It is important to note for our later work that this theorem does not
require $\mathbf{L}$ to be Laplacian or even positive semidefinite.

The condition $\gamma_{c}<\gamma_{c+1}$ implies a nonzero
eigengap~(\cite*{Chung1997}). In practice, the eigengap is often
used as a criterion to determine the number of classes in clustering
scenarios. An idealized situation is that the multiplicity of the
eigenvalue zero is $c$.

%%
%$\bolds{\Pi}$
%Problem~(\ref{eqrelax1})
%%
%%
%$\mathbf{U}' \mathbf{E}\mathbf{G}$ as
%$\mathbf{U}' \mathbf{E}\mathbf{G}= \bolds{\Theta}\bolds{\Lambda
%}\mathbf{V}'$ and let $\mathbf{Q}= \bolds{\Theta}\mathbf{V}'$
%%
%compute $t_i = \mathop{\arg\max}_{j} y_{ij}$, and recompute $\mathbf
%{E}$ by allocating
%the $i$th data point to class $t_i$ if $\max_{j} y_{ij} >0$ and
%to class $c$ otherwise
%%
%%

%%%%%%%%%%%%%%%%%%%%%%%%%%%%%%%%%%%%%%%%%%%%%%%%%%%%%%%%%%%%%%%%%%%%%%%%%%%%
%%%%%%%%%%%%%%%%%%%%%%%%%%%%%%%%%%%%%%%%%%%%%%%%%%%%%%%%%%%%%%%%%%%%%%%%%%%%
%s3 ###
\section{Rounding Schemes} \label{secrounding}

We now consider the problem of rounding---trans-\mbox{forming} the
real-valued solution of a spectral relaxation problem into
a discrete set of values that can be interpreted as a clustering.
In this section we present two different solutions to the
rounding problem, one based on Procrustes analysis and the
other based on the $K$-means algorithm.

%In the binary case ($c{=}2$), we have $\Q= 1$ or $-1$. Moreover,
%Let us now consider the rounding problem with the relaxation matrix
%$\Y$ obtained from the previous two sections.
%Writing ${\Y} =[{\y}_1, \ldots,
%{\y}_n]'=[{y}_{ij}]$, the rounding problem is to assign data points
%to classes based on $\Y$.

%%%%%%%%%%%%%%%%%%%%%%%%%%%%%%%%%%%%%%%%%%%%%%%%%%%%%%%%%%%%%%%%%%%%%%%%%%%%
%%%%%%%%%%%%%%%%%%%%%%%%%%%%%%%%%%%%%%%%%%%%%%%%%%%%%%%%%%%%%%%%%%%%%%%%%%%%
%s3.1 ###
\subsection{Procrustean Transformation for Rounding} \label{secprocr}

In Theorem~\ref{THMCEP0} we have shown that the solution of the
spectral relaxation problem is a matrix
$\mathbf{Y}= \bolds{\Pi}^{-1/2} \mathbf{U}\mathbf{Q}$,
where $\mathbf{Q}$ is an arbitrary orthogonal matrix. We have also
seen, in
Proposition~\ref{PRO1}, that a matrix $\mathbf{Y}$ in which the columns
of $\mathbf{Y}$ are piecewise constant with respect to a partition
$\mathbf{E}$
provides a representation of the objective function value \textsc{Pcut}.
If we had such a matrix $\mathbf{Y}$ in hand we could straightforwardly
find the partition $\mathbf{E}$: Letting $t_i=\mathop{\arg\max}_{j}\{y_{ij}\}$, allocate
$\mathbf{x}_i$ to the $t_i$th class if $y_{i t_i}>0$ and to the $c$th class
otherwise. On the other hand, if we had the partition we could
attempt to find an orthogonal matrix $\mathbf{Q}$ such that
$\mathbf{Y}= \bolds{\Pi}^{-1/2} \mathbf{U}\mathbf{Q}$ is
as close as possible to
the partition. This latter problem can be treated as a problem
in Procrustes analysis (\cite*{GowerBook2004}).

Specifically, given an indicator matrix $\mathbf{E}$ we pose the
following Procrustes problem:
%
%e3.1 ###
\begin{equation} \label{eqprocrust}
\hspace*{22pt}\mathop{\arg\min}_{\mathbf{Q}} L(\mathbf{Q}) = \operatorname
{tr}(\mathbf{E}\mathbf{G}- \mathbf{U}\mathbf{Q})(\mathbf{E}\mathbf
{G}- \mathbf{U}\mathbf{Q})',
\end{equation}
where $\mathbf{G}=[\mathbf{I}_{c - 1}-\frac{1}{c} \mathbf{1}_{c - 1}
\mathbf{1}_{c - 1}', -\frac{1}{c} \mathbf{1}_{c - 1}]'$. This problem has an analytical
solution: Denote the singular value decomposition of
$\mathbf{U}'\mathbf{E}\mathbf{G}$
as $\mathbf{U}' \mathbf{E}\mathbf{G}= \bolds{\Theta}\bolds
{\Lambda}\mathbf{V}'$. Then the minimizing value of $\mathbf{Q}$ in
$L$ is given by $\mathbf{Q}= \bolds{\Theta}\mathbf{V}'$
(see, e.g., \cite*{Mardia1979}, page 416).

We summarize this Procrustean approach to rounding in algorithmic
form in Algorithm~\ref{alg1} in the context of a generic spectral
clustering algorithm.

\def\figurename{\normalfont{\textbf{Algorithm}}}
\begin{figure}
\caption{\label{alg1} Spectral Clustering with Procrustean Rounding}
\begin{tabular*}{\columnwidth}{@{\extracolsep{\fill}}p{240pt}}
\hline
\hspace*{0.6mm}1:\hspace*{2mm}\textit{Input}: An affinity matrix $\mathbf{W}$ and a diagonal
\mbox{ma-}
\hspace*{6mm}trix~$\bolds{\Pi}$
\\
\hspace*{0.6mm}2:\hspace*{2mm}\textit{Relaxation}: Obtain $\mathbf{Y}= \bolds{\Pi}^{-1/2} \mathbf{U}\mathbf{Q}$
from \mbox{pro-}
\hspace*{6mm}blem~(\ref{eqrelax1})
\\
\hspace*{0.6mm}3:\hspace*{2mm}\textit{Initialize}: Choose the initial partition
$\mathbf{E}$
\\
\hspace*{0.6mm}4:\hspace*{2mm}\textit{Rounding}: Repeat the following procedure until
\hspace*{6mm}convergence:
\\
\hspace*{9mm}(a)\hspace*{2mm}Recompute $\mathbf{E}\mathbf{G}$, implement the SVD of
\hspace*{15.5mm}$\mathbf{U}'\mathbf{E}\mathbf{G}$ as
$\mathbf{U}' \mathbf{E}\mathbf{G}= \bolds{\Theta}\bolds{\Lambda}\mathbf{V}'$
and let $\mathbf{Q}= \bolds{\Theta}\mathbf{V}'$
\\
\hspace*{9mm}(b)\hspace*{2mm}Recompute $\mathbf{Y}=[y_{ij}] = \bolds{\Pi}^{-1/2} \mathbf{U}\mathbf{Q}$,
compute
\hspace*{15.5mm}$t_i = \mathop{\arg\max}_{j} y_{ij}$, and recompute $\mathbf{E}$ by
\mbox{allo-}
\\
\hspace*{15.5mm}cating the $i$th data point to class $t_i$ if
\hspace*{15.5mm}$\max_{j} y_{ij} >0$ and
to class $c$ otherwise
\\
\hspace*{0.6mm}5:\hspace*{2mm}Output $\{t_1, \ldots, t_n\}$.
\\
\hline
\end{tabular*}
\end{figure}

\citet{YuICCV2003} have presented a rounding algorithm that is
similar to the Procrustean approach we have presented but different
in detail. The authors work with an $n\times c$ matrix $Z$ and solve
the relaxation
$\min\operatorname{tr}(\mathbf{Z}' \mathbf{L}\mathbf{Z})$
subject to $\mathbf{Z}'\mathbf{D}\mathbf{Z}= \mathbf{I}_c$.
Given the solution $\mathbf{Z}$ of this relaxation, the authors then compute
$\hat{\mathbf{Z}}= [\hat{z}_{ij}]=\operatorname{dg}({\mathbf{Z}}{\mathbf{Z}}')^{-1/2} {\mathbf{Z}}$.
Their rounding scheme is to allocate the $i$th data point to
class $t_i$ if $t_i=\mathop{\arg\max}_{j}{\hat{z}_{ij}}$. This
method can
be viewed as imposing a constraint; in particular, note that
the norms of the rows of $\hat{\mathbf{Z}}$ are equal to 1.
To motivate this constraint, the authors assume that the
solution $\mathbf{Z}$ can be expressed as a rescaling of $\hat
{\mathbf{Z}}$:
$\mathbf{Z}= \hat{\mathbf{Z}}(\hat{\mathbf{Z}}' \mathbf{D}\hat
{\mathbf{Z}})^{-1/2}$. Inverting
this expression yields
$\hat{\mathbf{Z}}= \operatorname{dg}({\mathbf{Z}} {\mathbf{Z}}')^{-1/2}{\mathbf{Z}}$.
But it is not clear that a solution $\mathbf{Z}$ of the relaxation
can be expressed in this form; the constraints on $\hat{\mathbf{Z}}$
are not incorporated into the relaxation. The use of $\hat{\mathbf{Z}}$
defined in this way must be viewed as a heuristic post-processing
procedure. The Procrustean approach that we have presented
in this section provides a resolution of this difficulty;
that approach requires no post-processing of the matrix obtained
from the spectral relaxation.

We return to the Procrustean approach in Section~\ref{secmargin},
where we provide additional justification for Procrustean rounding
based on a connection to margin maximization.

%%%%%%%%%%%%%%%%%%%%%%%%%%%%%%%%%%%%%%%%%%%%%%%%%%%%%%%%%%%%%%%%%%%%%%
%s3.2 ###
\subsection{$K$-means for Rounding} \label{seckmeans}

Another approach to removing the ``nuisance'' orthogonal matrix
$\mathbf{Q}$ is to consider rounding methods that are invariant to
rotation. The standard $K$-means algorithm provides an example,
and numerous authors have proposed using $K$-means on the embedding
obtained from spectral relaxation as a heuristic rounding
procedure~(\cite*{LuxburgSC2007}). \citet{Bach2006} have made
this approach more formal by showing that (weighted) $K$-means
arises when the rounding problem is formalized in terms of a
difference between projection matrices. In this section we
review this formulation within our nonredundant representation
of spectral relaxation.

Let us rewrite \textsc{Pcut} as
\[
\mbox{\textsc{Pcut}} = \operatorname{tr}(\mathbf{E}' \mathbf{H}_{\pi}
\mathbf{L}\mathbf{H}_{\pi}' \mathbf{E}(\mathbf{E}'\bolds{\Pi}\mathbf{E})^{-1}),
\]
where we define
$\mathbf{H}_{\pi}=\mathbf{I}_{n}-\frac{1}{\bolds{\pi}^{\prime}\mathbf{1}_n}\bolds{\pi}^{\prime}\mathbf{1}_n$
where we use the fact that
$\mathbf{H}_{\pi} \mathbf{L}\mathbf{H}_{\pi}' = \mathbf{L}$.
Defining
$\mathbf{E}_{\pi} \triangleq\mathbf{H}_{\pi}'\cdot\break \mathbf{E}(\mathbf{E}'\bolds{\Pi}\mathbf{E})^{-1/2}$,
we observe that the number of degrees of freedom of both
$\mathbf{Y}$ and $\mathbf{E}_{\pi}$ is
$(n - 1)(c - 1)$. Moreover, given
that $\mathbf{E}_{\pi}' \bolds{\Pi}\mathbf{E}_{\pi}
= \mathbf{I}_c - (\mathbf{E}' \bolds{\Pi}\mathbf{E})^{-1/2}
\mathbf{E}'\bolds{\pi}\bolds{\pi}'\cdot\break \mathbf{E}(\mathbf{E}' \bolds{\Pi}\mathbf{E})^{-1/2}/
(\bolds{\pi}'\mathbf{1}_n)$ and
$\bolds{\pi}'\mathbf{E}(\mathbf{E}' \bolds{\Pi}\mathbf{E})^{-1} \mathbf{E}' \bolds{\pi}
= \bolds{\pi}'\mathbf{1}_n$, there exists a
$c\times c$ permutation matrix $\mathbf{P}$ such that
\[
\mathbf{P} \mathbf{E}_{\pi}' \bolds{\Pi}\mathbf{E}_{\pi} \mathbf{P}' =
\left[
\matrix{
\mathbf{I}_{c - 1} & \mathbf{0}\cr
\mathbf{0} & 0
}
\right] =\left[
\matrix{
\mathbf{Y}'
\cr \mathbf{0}
}
\right]
\bolds{\Pi}[\mathbf{Y}, \mathbf{0}];
\]
this suggests viewing $\mathbf{Y}$ as an approximation to $\mathbf
{E}_{\pi}$ in
the metric given by $\bolds{\Pi}$. We quantify this by defining the
following distortion between the projection matrices defined by
$\mathbf{Y}$ and $\mathbf{E}_{\pi}$:
\begin{eqnarray*}
J_k(\mathbf{E}_{\pi}, \mathbf{Y})
&=& \tfrac{1}{2} \|\mathbf{Y}\bolds{\Pi}\mathbf{Y}'- \mathbf{E}_{\pi} \bolds{\Pi}\mathbf{E}_{\pi}'
\|_{F}^2\\
&=& c - 1 - \operatorname{tr}(\mathbf{Y}' \bolds{\Pi}\mathbf{E}
(\mathbf{E}'\bolds{\Pi}\mathbf{E})^{-1} \mathbf{E}'\bolds{\Pi}\mathbf{Y}).
\end{eqnarray*}
This objective function can be represented as the solution
of a weighted $K$-means problem, as shown by the following
result which is due to \citet{Bach2006}:
\begin{theorem} \label{thmwkm}
Let $\mathbf{Y}=[\mathbf{y}_1, \ldots, \mathbf{y}_n]'$ be a solution of
problem~(\ref{eqrelax1}). For any partition
$\{V_1, \ldots, V_c\}$, the criterion $F(\mathbf{m}_1, \ldots,
\mathbf{m}_c)=\sum_{j=1}^c \sum_{i\in V_j} \|\mathbf{y}_i - \mathbf{m}_j \|^2$
achieves its minimum
$J_k(\mathbf{E}_{\pi}, \mathbf{Y})$ at $\mathbf{m}_j =
\frac{1}{\sum_{i \in V_j} \pi_i}\cdot
\break\sum_{i \in V_j} \pi_i \mathbf{y}_i$.
\end{theorem}

Thus by updating the mean vectors $\mathbf{m}_j$ in the weighted $K$-means
algorithm we match the criterion $J_k(\mathbf{E}_{\pi}, \mathbf{Y})$, and by updating
the partition using weighted $K$-means we go downhill in the criterion.

Note that in the special case of the \textsc{Rcut} formulation, we
obtain the conventional unweighted $K$-means algorithm (given that
$\pi_i=1$ in that case).

We summarize the $K$-means approach to rounding in algorithmic
form in Algorithm~\ref{alg2}.

\def\figurename{\normalfont{\textbf{Algorithm}}}
\begin{figure}
\caption{\label{alg2} Spectral Clustering with $K$-means Rounding}
\begin{tabular*}{\columnwidth}{@{\extracolsep{\fill}}p{240pt}}
\hline
\hspace*{0.6mm}1:\hspace*{2mm}\textit{Input}: An affinity matrix $\mathbf{W}$ and a diagonal
\mbox{ma-}
\\
\hspace*{6mm}trix~$\bolds{\Pi}$
\\
\hspace*{0.6mm}2:\hspace*{2mm}\textit{Relaxation}: Obtain $\mathbf{Y}= \bolds{\Pi}^{-1/2} \mathbf{U}\mathbf{Q}$ from
\mbox{problem}
\\
\hspace*{6mm}(\ref{eqrelax1})
\\
\hspace*{0.6mm}3:\hspace*{2mm}\textit{Initialize}: Choose the initial partition $\mathbf{E}$
\\
\hspace*{0.6mm}4:\hspace*{2mm}\textit{Rounding}: Repeat the following procedure until
\\
\hspace*{6mm}convergence:
\\
\hspace*{9mm}(a) Compute $\mathbf{m}_j = \frac{1}{\sum_{i \in V_j} \pi_i}\sum_{i \in V_j} \pi_i \mathbf{y}_i$
\\
\hspace*{9mm}(b) Find $t_i = \mathop{\arg\min}_{j} \|y_{i} - \mathbf{m}_j\|$,
and \mbox{recom-}
\\
\hspace*{14.5mm}pute~$\mathbf{E}$ by allocating the $i$th data point to
\\
\hspace*{14.5mm}class~$t_i$
\\
\hspace*{0.6mm}5:\hspace*{2mm}\textit{Output} $\{t_1, \ldots, t_n\}$.
\\
\hline
\end{tabular*}
\end{figure}

%%%%%%%%%%%%%%%%%%%%%%%%%%%%%%%%%%%%%%%%%%%%%%%%%%%%%%%%%%%%%%%%%%%%%%%%%%%%%
%%%%%%%%%%%%%%%%%%%%%%%%%%%%%%%%%%%%%%%%%%%%%%%%%%%%%%%%%%%%%%%%%%%%%%%%%%%%%%
%s4 ###
\section{Spectral Clustering and Minimum-Variance Criteria} \label{secvarrelax}

In this section and the following two sections we present some
relationships between spectral clustering and various topics
in statistics. Our goal is both to illuminate the spectral
approach and to suggest directions for further research.

Minimum-variance clustering is a classical approach to
clustering~(\cite*{Webb2002}). In this section, following
\citet{ZhaNIPS2002} and \citet{DhillonPAMI2007}, we present
spectral solutions to the minimum-variance clustering problem,
and we establish connections between minimum-variance clustering
and the \textsc{Pcut} framework.

Let $\{\mathbf{x}_1, \ldots, \mathbf{x}_n\} \in\mathcal{X}\subset\mathbb
{R}^d$ denote the
observed data. The pooled within-class covariance matrix $\mathbf{S}_W$
is given by
\[
\mathbf{S}_W = \frac{1}{n} \sum_{j=1}^c \sum_{i \in V_j}
(\mathbf{x}_i- \mathbf{m}_j) (\mathbf{x}_i- \mathbf{m}_j)',
\]
where ${\mathbf{m}}_j = \frac{1}{n_j} \sum_{i \in V_j} {\mathbf{x}}_i$. Consider
the trace of the within-class covariance matrix:
\[
\operatorname{tr}(\mathbf{S}_W) = \frac{1}{n} \sum_{j=1}^c
\sum_{i\in V_j} \|\mathbf{x}_i -\mathbf{m}_j\|^2.
\]
Clustering algorithms which are based on the minimization of this
trace are referred to as minimum-variance methods.

In order to establish a connection with the spectral relaxation
presented in Section~\ref{secpcutrelax}, we define a weighted pooled
within-class covariance matrix in an reproducing kernel Hilbert space
(RKHS) induced by a reproducing kernel $K$. In particular, assume
that we are given the reproducing kernel
$K\dvtx \mathcal{X}\times\mathcal{X}\rightarrow\mathbb{R}$ such
that
$K(\mathbf{x}_i,\mathbf{x}_j)$ $=$
$\bolds{\phi}(\mathbf{x}_i)'\bolds{\phi} (\mathbf{x}_j)$ for
$\mathbf{x}_i, \mathbf{x}_j\in\mathcal{X}$,
where
$\bolds{\phi}(\mathbf{x})$ is called a \textit{feature vector} corresponding to a data
point $\mathbf{x}\in\mathcal{X}$. In the sequel, we use the tilde
notation %$\ti$
to denote feature vectors. Thus, the data matrix in the feature
space is denoted as
$\tilde{\mathbf{X}}
=[\tilde{\mathbf{x}}_1, \tilde{\mathbf{x}}_2,\ldots, \tilde{\mathbf{x}}_n]'$.
The centered kernel matrix takes the
form $\mathbf{K}= \mathbf{H}_n \tilde{\mathbf{X}} \tilde{\mathbf{X}}' \mathbf{H}_n$; note that it
is positive semidefinite and satisfies $\mathbf{K}\mathbf{1}_n =
\mathbf{0}$.

Generalizing slightly, we introduce weighted versions of the
sample covariance matrix $\tilde{\mathbf{S}}$, the between-class
covariance matrix $\tilde{\mathbf{S}}_B$ and the within-class
covariance matrix
$\tilde{\mathbf{S}}_W$:
\begin{eqnarray*}
\tilde{\mathbf{S}} &=& \frac{1}{\sum_{i=1}^n \pi_i} \sum_{i=1}^n\pi_i
(\tilde{\mathbf{x}}_i - \tilde{\mathbf{m}})
(\tilde{\mathbf{x}}_i - \tilde{\mathbf{m}})', \\
\tilde{\mathbf{S}}_B &=& \frac{1}{\sum_{i=1}^n \pi_i} \sum_{j=1}^c\sum_{i\in V_j}
\pi_i (\tilde{\mathbf{m}}_j - \tilde{\mathbf{m}})
(\tilde{\mathbf{m}}_j - \tilde{\mathbf{m}})', \\
\tilde{\mathbf{S}}_W &=& \frac{1}{\sum_{i=1}^n \pi_i}
\sum_{j=1}^c\sum_{i\in V_j}
\pi_i (\tilde{\mathbf{x}}_i - \tilde{\mathbf{m}}_j)
(\tilde{\mathbf{x}}_i - \tilde{\mathbf{m}}_j)',
\end{eqnarray*}
where the $\pi_i$ are known positive weights, $\tilde
{\mathbf{m}} =
\frac{1}{\sum_{i=1}^n \pi_i }\cdot\break\sum_{i=1}^n\pi_i \tilde{\mathbf{x}}_i$
and
$\tilde{\mathbf{m}}_j = \frac{1}{\sum_{i \in V_j} \pi_i}\sum_{i\in V_j}
\pi_i\tilde{\mathbf{x}}_i$.
It is clear that $\tilde{\mathbf{S}}_W = \tilde{\mathbf{S}} - \tilde{\mathbf{S}}_B$.

We now formulate a minimum-variance clustering problem in the RKHS
as the minimization of $\operatorname{tr}(\tilde{\mathbf{S}}_W)$,
which is given by
\[
\operatorname{tr}(\tilde{\mathbf{S}}_W) = \frac{1}{\sum_{i=1}^n
\pi_i} \sum_{j=1}^c \sum_{i
\in V_j} \pi_i \|\tilde{\mathbf{x}}_i - \tilde{\mathbf{m}}_j \|^2.
\]
Like the minimization of \textsc{Pcut}, this minimization is computationally
infeasible in general. It is therefore natural to consider minimizing
$\operatorname{tr}(\tilde{\mathbf{S}}_W)$ by using the spectral
relaxations presented in
Section~\ref{secrelax1}. We present a way to do this in the following
section.

%s4.1 ###
\subsection{Spectral Relaxation in the RKHS}\label{secrelax2}

Let us rewrite $\tilde{\mathbf{S}}$ and $\tilde{\mathbf{S}}_{B}$ as
\[
\tilde{\mathbf{S}} = \frac{1}{\bolds{\pi}'\mathbf{1}_n}
\tilde{\mathbf{X}}' \mathbf{H}_{\pi} \bolds{\Pi}\mathbf{H}_{\pi}'
\tilde{\mathbf{X}}
\]
and
\[
\tilde{\mathbf{S}}_{B} = \frac{1}{\bolds{\pi}' \mathbf{1}_n}
\tilde{\mathbf{X}}' \mathbf{H}_{\pi} \bolds{\Pi}\mathbf{E}
\bigl(\mathbf{E}' \bolds{\Pi}\mathbf{E}\bigr)^{-1} \mathbf{E}'
\bolds{\Pi}\mathbf{H}_{\pi}' \tilde{\mathbf{X}},
\]
recalling that $\mathbf{H}_{\pi} =
\mathbf{I}_{n} -
\frac{1}{\bolds{\pi}'\mathbf{1}_n}
\bolds{\pi}\mathbf{1}_n'$. This yields
\begin{eqnarray*}
&&\tilde{\mathbf{S}}_W = \frac{1}{\bolds\pi' \mathbf{1}_{n}}
\bigl[\tilde{\mathbf{X}}' \mathbf{H}_{\pi} \bolds{\Pi}
\mathbf{H}_{\pi}' \tilde{\mathbf{X}}
\\
&&\hspace*{54pt}{}- \tilde{\mathbf{X}}'
\mathbf{H}_{\pi} \bolds{\Pi}\mathbf{E}
\bigl(\mathbf{E}' \bolds{\Pi}\mathbf{E}\bigr)^{-1}
\mathbf{E}'\bolds{\Pi}\mathbf{H}_{\pi}' \tilde{\mathbf{X}} ].
\end{eqnarray*}
The minimization of $\operatorname{tr}(\tilde{\mathbf{S}}_W)$ is
thus equivalent to the
maximization of
%
%e4.1 ###
\begin{equation} \label{eqssp}
T = \operatorname{tr}(\mathbf{E}' \bolds{\Pi}\mathbf{H}_{\pi}'
\mathbf{K}\mathbf{H}_{\pi} \bolds{\Pi}\mathbf{E}
(\mathbf{E}' \bolds{\Pi}\mathbf{E})^{-1} ),
\end{equation}
because
$\tilde{\mathbf{X}}' \mathbf{H}_{\pi} \bolds{\Pi}\mathbf{H}_{\pi}' \tilde{\mathbf{X}}$ is
independent of
$\mathbf{E}$ and we have
$\mathbf{H}_n \mathbf{H}_{\pi} = \mathbf{H}_{\pi}$.
Let
$\bolds{\Delta}=[\delta_{ij}^2]$,
where $\delta_{ij}$ is the squared distance between $\tilde{\mathbf{x}}_i$
and $\tilde{\mathbf{x}}_j$, that is,
\begin{eqnarray*}
\delta_{ij}^2
&=& (\tilde{\mathbf{x}}_i-\tilde{\mathbf{x}}_j)'(\tilde{\mathbf{x}}_i - \tilde{\mathbf{x}}_j)'
\\
&=& K(\mathbf{x}_i, \mathbf{x}_i) + K(\mathbf{x}_j, \mathbf{x}_j) - 2 K(\mathbf{x}_i, \mathbf{x}_j).
\end{eqnarray*}
Given that
$-\frac{1}{2} \mathbf{H}_{\pi}' \bolds{\Delta}\mathbf{H}_{\pi}
= \mathbf{H}_{\pi}' \mathbf{K}\mathbf{H}_{\pi}$,
the minimization of $\operatorname{tr}(\tilde{\mathbf{S}}_W)$
is thus equivalent to that of
$\operatorname{tr}(\mathbf{E}'\bolds{\Pi}\cdot
\break\mathbf{H}_{\pi}'
\bolds{\Delta}\mathbf{H}_{\pi}\bolds{\Pi}\mathbf{E}(\mathbf{E}' \bolds{\Pi}\mathbf{E})^{-1} )$.

Recall that in the proof of Proposition~\ref{PRO1}, $\mathbf{L}$ is
required to satisfy only the conditions $\mathbf{L}=\mathbf{L}'$ and
$\mathbf{L}\mathbf{1}_n =~\mathbf{0}$.
Note that $\bolds{\Pi}\mathbf{H}_{\pi}' \mathbf{K}\mathbf{H}_{\pi}
\bolds{\Pi}\mathbf{1}_n = \mathbf{0}$. Thus, if $\mathbf{Y}$
is an $n\times (c - 1)$ matrix subject to the three conditions in
Proposition~\ref{PRO1}, we have $T = \operatorname{tr}(\mathbf{Y}'
\bolds{\Pi}\mathbf{H}_{\pi}' \mathbf{K}\mathbf{H}_{\pi}\cdot
\break \bolds
{\Pi}\mathbf{Y})$. This allows us to relax the
maximization of $T$ with respect to $\mathbf{E}$ as follows:
%
%e4.2 ###
\begin{eqnarray} \label{eqrelax5}
&&\max_{\mathbf{Y}\in\mathbb{R}^{n\times (c - 1)}}
\operatorname{tr}(\mathbf{Y}'\bolds{\Pi}\mathbf{H}_{\pi}'\mathbf{K}\mathbf{H}_{\pi} \bolds{\Pi}{\mathbf{Y}})
\nonumber
\\
&&\quad=\operatorname{tr}(\mathbf{Y}'\bolds{\Pi}\mathbf{K}\bolds{\Pi}\mathbf{Y})
\\
&&\quad\mbox{ s.t. }\mathbf{Y}' \bolds{\Pi}\mathbf{Y}= \mathbf{I}_{c - 1}\mbox{ and }
\mathbf{Y}' \bolds{\Pi}\mathbf{1}_n = \mathbf{0},\nonumber
\end{eqnarray}
where the second equality in the objective is due to the identity
$\mathbf{Y}'\bolds{\Pi}\mathbf{H}_{\pi}'
= \mathbf{Y}' \bolds{\Pi}$.
Letting $\mathbf{Y}_0 = \bolds{\Pi}^{1/2} \mathbf{Y}$ leads to
%
%e4.3 ###
\begin{eqnarray} \label{EQRELAX6}
\nonumber&&\max_{\mathbf{Y}_0 \in\mathbb{R}^{n\times (c - 1)}}
\operatorname{tr}(\mathbf{Y}_0'\bolds{\Pi}^{1/2} \mathbf{H}_{\pi}' \mathbf{K}\mathbf{H}_{\pi}
\bolds{\Pi}^{1/2} \mathbf{Y}_0)
\\[-8pt]\\[-8pt]
\nonumber&&\quad\mbox{s.t. }\mathbf{Y}_0' \mathbf{Y}_0 = \mathbf{I}_{c - 1}
\mbox{ and }
\mathbf{Y}_0'\bolds{\Pi}^{1/2} \mathbf{1}_n = \mathbf{0}.
\end{eqnarray}
This optimization problem is solved in Appendix~\ref{aprkhs}.
In particular, let
$\mathbf{U}$ be an $n\times (c - 1)$ matrix whose columns are the top $c - 1$ eigenvectors of
$\bolds{\Pi}^{1/2} \mathbf{H}_{\pi}' \mathbf{K}\cdot\break\mathbf{H}_{\pi}\bolds{\Pi}^{1/2}$.
The solution of problem~(\ref{EQRELAX6}) is then $\mathbf{Y}_0 =
\mathbf{U}\mathbf{Q}$
where $\mathbf{Q}$ is an arbitrary $(c - 1)\times (c - 1)$ orthonormal
matrix. Hence, the solution of problem (\ref{eqrelax5}) is
$\mathbf{Y}= \bolds{\Pi}^{-1/2} \mathbf{U}\mathbf{Q}$.

%%%%%%%%%%%%%%%%%%%%%%%%%%%%%%%%%%%%%%%%%%%%%%%%%%%%%%%%%%%%%%%%%%%%%%%%%%%%%
%%%%%%%%%%%%%%%%%%%%%%%%%%%%%%%%%%%%%%%%%%%%%%%%%%%%%%%%%%%%%%%%%%%%%%%%%%%%%
%s4.2 ###
\subsection{Minimum Variance Formulations versus \textsc{Pcut} Formulations}\label{secpcutmiv}

Since the Laplacian matrix $\mathbf{L}$ is symmetric and positive semidefinite,
its Moore--Penrose (MP) inverse is also positive semidefinite.
Thus we can regard $\mathbf{L}$ as the MP inverse of a kernel matrix
$\mathbf{K}$
and investigate the relationship between the spectral relaxations
obtained from the minimum variance and the \textsc{Pcut} formulations.
In fact, we have the following theorem, whose proof is given
in Appendix~\ref{apmp}.

\begin{theorem} \label{THM6} Assume that $\mathbf{L}^{+}=\mathbf{K}$.
If $\operatorname{rk}(\mathbf{L})=\operatorname{rk}(\mathbf
{K})=n - 1$, then $\mathbf{Y}$ is the solution of
problem~(\ref{eqrelax1}) if and only if it is the solution of
problem~(\ref{eqrelax5}).
\end{theorem}

Thus, an equivalent formulation of spectral clustering based on
the \textsc{Pcut} criterion is obtained by considering the minimum variance
criterion with $\mathbf{K}= \mathbf{L}^{+}$. Note that $\bolds{\Pi
}$ consists of the
diagonal elements of $\mathbf{K}^{+}$ in the \textsc{Ncut} setting, so it
is not
expedient computationally to obtain $\bolds{\Pi}$ from $\mathbf
{K}$---we would need
to calculate $\mathbf{K}^{+}$. We thus suggest defining $\bolds{\Pi}=\mathbf{I}_n$ in the
minimum-variance setting, corresponding to the ratio cut formulation.

It is also possible to start from a minimum-variance formulation
(with $\bolds{\Pi}=\mathbf{I}_n$) and obtain a \textsc{Rcut} problem.
However, in the
corresponding \textsc{Rcut} problem, the matrix $\mathbf{K}^{+}$ is not guaranteed
to be Laplacian, because the off-diagonal entries of $\mathbf{K}^{+}$ are
possibly positive for an arbitrary kernel matrix $\mathbf{K}$. In this
case, we
can let
$\mathbf{L}= \mathbf{K}^{+} + n \beta\mathbf{H}_n$
where
$\beta= \min\{
\max_{i\neq j}\{ [\mathbf{K}^{+}]_{ij} \}, 0 \}$.
Such an $\mathbf{L}$ is Laplacian. Moreover, we have
$\operatorname{tr}(\mathbf{Y}'(\mathbf{K}^{+} + n\beta\mathbf{H}_n)\mathbf{Y})
=\break\operatorname{tr}(\mathbf{Y}'\mathbf{K}^{+} \mathbf{Y}) + n (c - 1) \beta$
due to
$\mathbf{Y}' \mathbf{Y}= \mathbf{I}_{c - 1}$
and $\mathbf{Y}'\cdot\break\mathbf{1}_n =~\mathbf{0}$.
Since $\min(\operatorname{tr}(\mathbf{Y}'(\mathbf{K}^{+} + n \beta\mathbf{H}_n)\mathbf{Y}))$
is equivalent to
$\min(\operatorname{tr}(\mathbf{Y}' \mathbf{K}^{+} \mathbf{Y}))$,
it is not necessary to compute the value of $\beta$.

It is worth noting that the condition
$\operatorname{rk}(\mathbf{L})=\break \operatorname{rk}(\mathbf{K})= n - 1$
is necessary. Without this condition,
$\bolds{\Pi}^{-1/2}\mathbf{L}\bolds{\Pi}^{-1/2}$
is a generalized inverse of
$\bolds{\Pi}^{1/2} \mathbf{H}_{\pi}' {\mathbf{L}^{+}}\break\mathbf{H}_{\pi}\bolds{\Pi}^{1/2}$,
because
\begin{eqnarray*}
&&\bolds{\Pi}^{1/2} \mathbf{H}_{\pi}' {\mathbf{L}^{+}}
\mathbf{H}_{\pi} \bolds{\Pi}^{1/2}\bolds{\Pi}^{-1/2}
\mathbf{L}\bolds{\Pi}^{-1/2}\bolds{\Pi}^{1/2}
\mathbf{H}_{\pi}' {\mathbf{L}^{+}}
\\
&&\quad{}\cdot\mathbf{H}_{\pi} \bolds{\Pi}^{1/2}
= \bolds{\Pi}^{1/2}\mathbf{H}_{\pi}'
{\mathbf{L}^{+}} \mathbf{H}_{\pi} \bolds{\Pi}^{1/2},
\end{eqnarray*}
but it is not necessarily the MP inverse. In this case, it is no
longer the case that
$\bolds{\Pi}^{-1/2} \mathbf{L}\bolds{\Pi}^{-1/2}$ and $\bolds{\Pi}^{1/2}
\mathbf{H}_{\pi}' {\mathbf{L}^{+}} \mathbf{H}_{\pi}\bolds{\Pi}^{1/2}$ are guaranteed to
have the same eigenvectors associated with nonzero eigenvalues.
Thus, in this case, even if $\mathbf{K}=\mathbf{L}^{+}$, the
solutions of (\ref{eqrelax5})
and (\ref{eqrelax1}) are different. In summary we see that the spectral
clustering formulations based on the minimum-variance criteria and
\textsc{Pcut}, while closely related, are not fully equivalent.

\citet{DhillonPAMI2007} pursue a slightly different connection
between minimum-variance criteria and spectral relaxation. They
formulate the minimum-variance criterion via the maximization of
%
%e4.4 ###
\begin{eqnarray} \label{eqobj}
T^{\prime} = \operatorname{tr}(\mathbf{E}' \bolds{\Pi}\mathbf
{K}\bolds{\Pi}\mathbf{E}(\mathbf{E}' \bolds{\Pi}\mathbf{E})^{-1}),
\end{eqnarray}
which is readily shown to be equal to $T +{\bolds\pi}' \mathbf{K}\cdot\break\bolds{\pi}/
({\bolds\pi}' \mathbf{1}_n)$,
where $T$ is defined by (\ref{eqssp}). Thus the maximization
of $T^{\prime}$ is equivalent to the maximization of $T$.
\citet{DhillonPAMI2007} then formulate the cut minimization
problem as an equivalent maximization problem:
\[
\max\bigl(\mathbf{E}' \bolds{\Pi}(\bolds{\Pi}^{-1} - \bolds{\Pi}^{-1}
\mathbf{L}\bolds{\Pi}^{-1})
\bolds{\Pi}\mathbf{E}(\mathbf{E}' \bolds{\Pi}\mathbf{E})^{-1}
\bigr),
\]
and treat
$\bolds{\Pi}^{-1}-\bolds{\Pi}^{-1} \mathbf{L}\bolds{\Pi}^{-1}$ as $\mathbf{K}$
in $T^{\prime}$. However, $\bolds{\Pi}^{-1} - \bolds{\Pi}^{-1}\mathbf{L}\bolds{\Pi}^{-1}$
is generally indefinite, a difficulty that the authors
circumvent by letting $\mathbf{K}= \rho\mathbf{I}_n - \mathbf{L}$
in \textsc{Rcut} and
$\mathbf{K}= \rho\mathbf{D}^{-1} + \mathbf{D}^{-1} \mathbf
{W}\mathbf{D}^{-1}$ in \textsc{Ncut}, where
$\rho$ is a constant chosen to make $\mathbf{K}$ positive semidefinite.

The idea of considering a kernel matrix that is the MP inverse
of a Laplacian matrix will return in later sections, in particular
in Section~\ref{sechyperplane} where we will see that it allows
us to provide a geometrical interpretation for spectral clustering,
and in Section~\ref{secgias}, where we present a probabilistic
interpretation of spectral relaxation.

\section{Spectral Clustering: A Margin-Based Perspective} \label{secmargin}

In this section we consider a margin-based perspective on
spectral clustering. First, we show that the margin-based
perspective provides us with insight into the relationship
between spectral embedding and rounding. In particular, we
show that the problems in (\ref{eqrelax1}) and (\ref{eqrelax5})
can be understood in terms of the fitting of hyperplanes in
an RKHS. For a data point $\mathbf{x}$, we show that the elements
of the embedding $\mathbf{y}$ are proportional to the signed distances
of feature vector $\tilde{\mathbf{x}}$ to each of these hyperplanes.
This provides support for the Procrustean rounding in which
rounding is achieved by nonmaximum suppression of the elements
of $\mathbf{y}$. Second, we provide some additional direct justification
for the Procrustean approach, showing that the rounding problem
can be analyzed in terms of the approximation of a margin-based
multiway classification criterion.

%%%%%%%%%%%%%%%%%%%%%%%%%%%%%%%%%%%%%%%%%%%%%%%%%%%%%%%%%%%%%%%%%%%%%%%
%%%%%%%%%%%%%%%%%%%%%%%%%%%%%%%%%%%%%%%%%%%%%%%%%%%%%%%%%%%%%%%%%%%%%%
%s5.1 ###
\subsection{Hyperplanes in the RKHS}\label{sechyperplane}

Let us consider a multiway classification problem. That is, we
consider a problem in which data points are pairs, $({\mathbf{x}}_i, t_i)$,
where $t_i$ is the label of the $i$th data point. Using the same
notation as in Section~\ref{secvarrelax}, the multiway classification
problem has the following standard formulation in an RKHS based on a
kernel function $K$:
%
%e5.1 ###
\begin{equation} \label{eqkerclass}
\min_{\bolds{\beta}_0, \mathbf{B}}\operatorname{tr}(\mathbf{B}'
\mathbf{K}\mathbf{B}) + \frac{\gamma}{n}\sum_{i=1}^n f_{t_i}(\mathbf{B}' \mathbf{k}_i + \bolds{\beta}_0),
\end{equation}
where $f_j(\cdot)$ is a convex surrogate of the 0--1 loss,
$\mathbf{k}_i=(K(\mathbf{x}_1, \mathbf{x}_i), \ldots, K(\mathbf{x}_n, \mathbf{x}_i))'$ is the $i$th column
of the kernel matrix $\mathbf{K}$, $\mathbf{B}=[\mathbf{b}_1, \ldots,
\mathbf{b}_{c - 1}]$
is an
$n\times (c - 1)$ matrix of regression vectors, $\bolds{\beta}_0$
is a
$(c - 1)\times 1$ vector of intercepts and $\gamma>0$ is a
regularization parameter.
We can use this optimization problem as the basis of a clustering
formulation by simply omitting the term
$\frac{\gamma}{n} \sum_{i=1}^n f_{t_i}(\cdot)$,
reflecting the fact that we have no labeled data in the
clustering setting. We obtain
%
%e5.2 ###
\begin{eqnarray}\label{eqkerclass1}
&&\nonumber\hspace*{6pt}\min_{\mathbf{B}} \operatorname{tr}(\mathbf{B}' \mathbf
{K}\mathbf{B}) \\[-8pt]\\[-8pt]
&&\nonumber\quad\mbox{s.t. }\mathbf{B}' \mathbf{K}\bolds{\Pi}\mathbf{1}_n =
\mathbf{0} \mbox{ and }\mathbf{B}' \mathbf{K}\bolds{\Pi}\mathbf{K}\mathbf{B}=\mathbf{I}_{c - 1}.
\end{eqnarray}

\def\figurename{\normalfont{\textsc{Fig.}}}
\setcounter{figure}{0}
\begin{figure*}[t]

\includegraphics{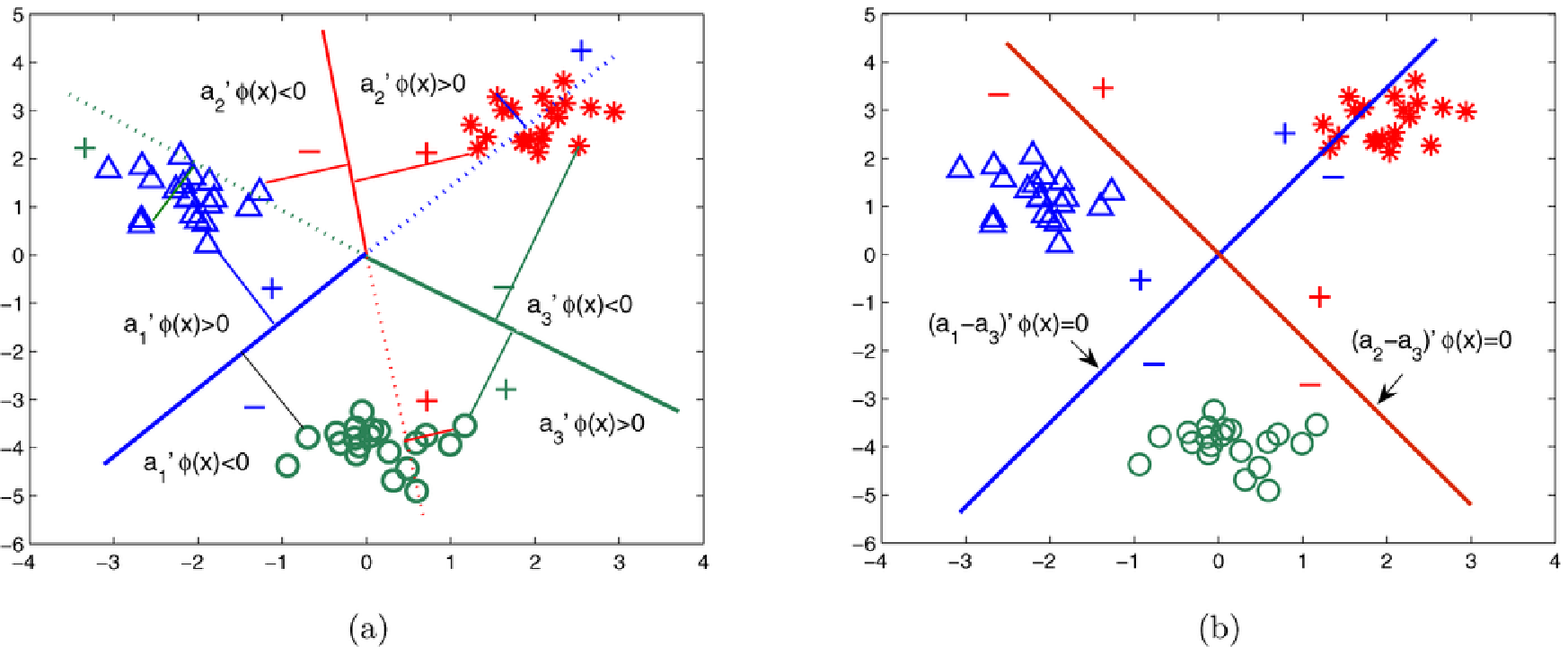}

  \caption{\label{fighyperplane} Illustrations of spectral clustering in the feature
space for a three-class separable example. The clustering
is based on the signed distances of the feature vector
$\tilde{\mathbf{x}}=\phi(\mathbf{x})$ to suitably defined hyperplanes.
\textup{(a)}
Hyperplanes in the feature space are represented by
their normals, $\mathbf{a}_j$, $j=1, 2, 3$, subject to the
sum-to-zero constraints. These hyperplanes are computed
from the vectors
$\mathbf{s}_1$ and $\mathbf{s}_2$ obtained from spectral
relaxation via
$\mathbf{a}_1= \mathbf{s}_1 - \frac{1}{3} (\mathbf{s}_1 + \mathbf{s}_2)$,
$\mathbf{a}_2 = \mathbf{s}_2 - \frac{1}{3} (\mathbf{s}_1 + \mathbf{s}_2)$
and
$\mathbf{a}_3= -\frac{1}{3} (\mathbf{s}_1 + \mathbf{s}_2)$.
\textup{(b)}
The hyperplanes
defined by the vectors $\mathbf{s}_1$ and $\mathbf{s}_2$.
Note that
$\mathbf{s}_1=\mathbf{a}_1 -\mathbf{a}_3$ and $\mathbf{s}_2=\mathbf{a}_2
- \mathbf{a}_3$.}
\end{figure*}

We now consider problem (\ref{eqkerclass1}) from two points of
view. From the first point of view, we let $\mathbf{Y}= \mathbf
{K}\mathbf{B}$ and
transform (\ref{eqkerclass1}) into
%
%e5.4 ###
%e5.3 ###
\begin{eqnarray}\label{eqkerclass2}
&&\nonumber\min_{\mathbf{Y}}\operatorname{tr}(\mathbf{Y}' \mathbf{K}^{+} \mathbf{Y})
\\[-8pt]\\[-8pt]
&&\nonumber\quad\mbox{s.t. }\mathbf{Y}' \bolds{\Pi}\mathbf{1}_n = \mathbf{0}
\mbox{ and }\mathbf{Y}' \bolds{\Pi}\mathbf{Y}= \mathbf{I}_{c - 1},
\end{eqnarray}
where we have used the identity $\mathbf{K}= \mathbf{K}\mathbf
{K}^{+} \mathbf{K}$. It is readily
seen that~(\ref{eqkerclass2}), and hence~(\ref{eqkerclass1}), is identical
with the spectral relaxation in~(\ref{eqrelax1}) by taking
$\mathbf{K}^{+}=\mathbf{L}$. We also obtain a relationship between~(\ref{eqkerclass2})
and~(\ref{eqrelax5}) from Section~\ref{secpcutmiv}; in particular,
in the special case in which $\operatorname{rk}(\mathbf{K})=n - 1$,
it follows from
Theorem~\ref{THM6} that~(\ref{eqkerclass2}) and~(\ref{eqrelax5})
are equivalent.

From a second point of view, we let
$\mathbf{S}= \tilde{\mathbf{X}}'\mathbf{B}$
(recall that $\tilde{\mathbf{X}}$ is the data matrix in the feature space).
The problem~(\ref{eqkerclass1}) is then transformed into
%
%e5.6 ###
%e5.5 ###
\begin{eqnarray}\label{eqkerclass3}
&&\nonumber\min_{\mathbf{S}}\operatorname{tr}(\mathbf{S}' \mathbf{S})
\\[-8pt]\\[-8pt]
&&\nonumber\quad\mbox{s.t. }\mathbf{S}' \tilde{\mathbf{X}}' \bolds{\Pi}\mathbf{1}_n
= \mathbf{0}\mbox{ and }\mathbf{S}' \tilde{\mathbf{X}}'
\bolds{\Pi}\tilde{\mathbf{X}} \mathbf{S}= \mathbf{I}
_{c - 1}.
\end{eqnarray}
Letting $\mathbf{S}=[\mathbf{s}_1, \ldots, \mathbf{s}_{c - 1}]$
denote the solution of
(\ref{eqkerclass3}), the equations
$\mathbf{s}_j' \tilde{\mathbf{x}}=0$, $j=1, \ldots, c-1$, define hyperplanes that pass through the weighted
centroid $\sum_{i=1}^n \pi_i \tilde{\mathbf{x}}_i$ of the feature vectors
$\tilde{\mathbf{x}}_i$. Moreover, the signed distance between feature vector
$\tilde{\mathbf{x}}_i$ and the hyperplane $\mathbf{s}_j' \tilde{\mathbf{x}}=0$ is
$\mathbf{s}_j'
\tilde{\mathbf{x}}_i$. Recall that
$\mathbf{Y}=[y_{ij}]= \mathbf{K}\mathbf{B}=\tilde{\mathbf{X}} \tilde{\mathbf{X}}'
\mathbf{B}= \tilde{\mathbf{X}} \mathbf{S}$. We thus have
$y_{ij}= \mathbf{s}_j' \tilde{\mathbf{x}}_i$. That is, $y_{ij}$ is the signed
distance of $\tilde{\mathbf{x}}_i$ to the $j$th hyperplane. We can therefore
interpret the spectral relaxation in (\ref{eqrelax1})~and (\ref{eqrelax5}) as yielding vectors whose elements are---using
the language of multiway classi\-fication---margin vectors.
Given this interpretation, it is reasonable to allocate labels by
finding the maximum element of $(y_{i1}, \ldots, y_{i,c{-}1}, 0)$.
This motivates the Procrustean approach to rounding, which can
be viewed as identifying boundaries between clusters by projecting
feature vectors onto hyperplanes in an RKHS. A graphical interpretation
of this result is provided in Figure~\ref{fighyperplane}.

%%
%
%(a) &
%(b) \\
%%
%space for a three-class separable example. The clustering
%is based on the signed distances of the feature vector
%$\tilde{\mathbf{x}}=\phi(\mathbf{x})$ to suitably-defined hyperplanes.
%(a) Hyperplanes in the feature space are represented by
%their normals, $\mathbf{a}_j$, $j=1, 2, 3$, subject to the
%sum-to-zero constraints. These hyperplanes are computed
%from the vectors $\mathbf{s}_1$ and $\mathbf{s}_2$ obtained from spectral
%relaxation via $\mathbf{a}_1= \mathbf{s}_1 -\frac{1}{3} (\mathbf{s}_1
%{+} \mathbf{s}_2)$,
%$\mathbf{a}_2 = \mathbf{s}_2 - \frac{1}{3} (\mathbf{s}_1 {+} \mathbf{s}_2)$,
%and $\mathbf{a}_3= -\frac{1}{3} (\mathbf{s}_1 {+} \mathbf{s}_2)$. (b)
%The hyperplanes
%defined by the vectors $\mathbf{s}_1$ and $\mathbf{s}_2$. Note that
%$\mathbf{s}_1=\mathbf{a}_1 {-}\mathbf{a}_3$ and $\mathbf{s}_2=\mathbf{a}_2
%{-}\mathbf{a}_3$.}
%
%%

%%%%%%%%%%%%%%%%%%%%%%%%%%%%%%%%%%%%%%%%%%%%%%%%%%%%%%%%%%%%%%%%%%%%%%%%%%%
%%%%%%%%%%%%%%%%%%%%%%%%%%%%%%%%%%%%%%%%%%%%%%%%%%%
%s5.2 ###
\subsection{Margin-Based Rounding Scheme} \label{secbayes}

We can also provide a direct connection between classification
and rounding. Let us return to the objective function in~(\ref{eqkerclass}), which we rewrite as
\[
\min_{\mathbf{Y}}\operatorname{tr}(\mathbf{Y}' \mathbf{K}^{+}
\mathbf{Y}) + \frac{\gamma}{n} \sum_{i=1}^n f_{t_i}(\mathbf{y}_i)
\]
by letting $\mathbf{Y}= \mathbf{K}\mathbf{B}$ and setting $\bolds
{\beta}_0=0$.
Assume that we have obtained a matrix $\mathbf{Y}$ from
spectral relaxation and recall that $\mathbf{Y}$ depends on an
arbitrary orthogonal matrix~$\mathbf{Q}$. From the classification
perspective we can view the subsequent rounding problem
as the problem of minimizing the classification loss
$\frac{1}{n}\sum_{i=1}^n f_{t_i}(\mathbf{y}_i)$ under the
constraint $\mathbf{Q}\mathbf{Q}' =\mathbf{I}_{c - 1}$. In this
section we
explore some of the consequences of this perspective.

In the multiway classification problem, we define
\textit{class-conditional probabilities} $P_j(\mathbf{x})$
for the $c$ classes $j=1,\ldots, c$. Using this
notation, we define the expected error at $\mathbf{x}$ as
follows:
%
%e5.7 ###
\begin{equation} \label{eqexperror}
R(\mathbf{x}, \mathbf{y})=\sum_{j=1}^c \mathbb{I}_{[t \neq j]} P_j(\mathbf{x}),
\end{equation}
where $t=\mathop{\arg\max}_{j} y_j$ or $t=c$ if $\max\{y_j\}<0$ and
where $\mathbb{I}_{[\#]}$ defines the 0--1 loss: it is 1 if $\#$ is
true and 0 otherwise. Since $\mathbb{I}_{[\bolds{\cdot}]}$ is a non-convex
objective function that leads to an intractable optimization
problem, the standard practice in the classification literature
is to replace $\mathbb{I}_{[\bolds{\cdot}]}$ with a ``surrogate loss
function'' $f_j(\mathbf{y})$ that is an upper bound on the 0--1
loss~(\cite*{BartlettJordanMcAuliffe2006}; \cite*{ShenWang07}).

The surrogate loss function that we consider in the current
paper is the following exponential loss:
%
%e5.8 ###
\begin{equation} \label{eqlosss}
f_j(\mathbf{y}) = \sum_{l \neq j}\exp(y_{l} - y_{j}),
\end{equation}
where for convenience we extend
$\mathbf{y}$ to a $c$-dimensional
vector in which
$y_{c} =0$. Note that the variables to
be optimized are the entries of the matrix $\mathbf{Q}$. Clearly,
$f_j(\mathbf{y})$ is an upper bound of $\mathbb{I}_{[t \neq j]}$,
because if $\mathbf{x}$ does not belong to class $j$, there exists at least
one $y_l$ such that $l \neq j$ and $y_l - y_j \geq0$, and hence
$\exp(y_{l} - y_{j}) \geq1$. This surrogate loss function also
has an important Fisher consistency property:

\begin{proposition} \label{THM5}
Assume $P_j(\mathbf{x})>0$ for $j=1,\break\ldots, c$. We then have
\begin{eqnarray*}
\hat{y}_j
&=&\mathop{\arg\max}_{\mathbf{y}} \sum_{j=1}^c \sum_{l \neq j} \exp(y_l - y_j)
P_j(\mathbf{x})
\\
&=& \frac{1}{2} \log\frac{P_j(\mathbf{x})}{P_c(\mathbf{x})}.
\end{eqnarray*}
\end{proposition}
The proof of Proposition~\ref{THM5} is a straightforward
calculation, so we omit it. This proposition shows that the surrogate
loss function that we have chosen is justified from the point of
view of classification as yielding a Bayes consistent
rule~(\cite*{BartlettJordanMcAuliffe2006}; \cite*{ZouZhuHastieTR2006}).

Returning to the rounding problem, we now consider the labels
$\{t_i\}$ as temporarily fixed and consider the empirical risk
function defined over the set of pairs $(\mathbf{x}_i, t_i)$ given by
\[
J(\mathbf{Q}) = \frac{1}{n} \sum_{i=1}^n \sum_{l \neq t_i}
\exp(y_{i l} - y_{i t_i}).
\]
We wish to optimize this empirical risk with respect to $\mathbf{Q}$.
This problem does not have a closed-form solution under the constraint
$\mathbf{Q}\mathbf{Q}'=\mathbf{I}_{c - 1}$. However, we can consider
a Taylor expansion
around $y_{ij}=0$. We have
\[
J(\mathbf{Q}) \approx(c - 1) - \frac{c}{n}
\sum_{i=1}^n \mathbf{g}_{t_i}' {\mathbf{y}}_i
+ c^2 \sum_{i=1}^n \pi_i^{-1},
\]
where $\mathbf{g}_j$ is the $j$th column of
$\mathbf{G}' =[\mathbf{I}_{c - 1}-\frac{1}{c} \mathbf{1}_{c - 1}
\mathbf{1}_{c - 1}',\break -\frac{1}{c} \mathbf{1}_{c - 1}]$, and where we have used the
fact that
${\mathbf{y}}_i' {\mathbf{g}_{t_i}} \cdot\break\mathbf{g}_{t_i}'
{\mathbf{y}}_i = \pi_i^{-1} \bolds{\mu}_i' \mathbf{Q}{\mathbf{g}_{t_i}}
\mathbf{g}_{t_i}' \mathbf{Q}' \bolds{\mu}_i \leq1/\pi_i$ because
$\mathbf{I}_{c - 1}-\break {\mathbf{g}_{t_i}} \mathbf{g}_{t_i}'$ is positive
semidefinite.
We thus see that the maximization of the linear term
$\sum_{i=1}^n \mathbf{g}_{t_i}' {\mathbf{y}}_i$ with respect to $\mathbf{Q}$
yields an approximate procedure for minimizing $J(\mathbf{Q})$.
But this is precisely the Procrustean problem~(\ref{eqprocrust})
discussed in Section~\ref{secrounding}.

It would also be possible to attempt to optimize $J(Q)$ directly
by making use of Newton or conjugate gradient methods on the
Stiefel manifold~(\cite*{EdelmanSIAM1999}).

%%%%%%%%%%%%%%%%%%%%%%%%%%%%%%%%%%%%%%%%%%%%%%%%%%%%%%%%%%%%%%%%%%%%%%%
%s6 ###
\section{Spectral Relaxation: The View From Gaussian Intrinsic Autoregression}\label{secgias}
%%%%%%%%%%%%%%%%%%%%%%%%%%%%%%%%%%%%%%%%%%%%%%%%%%%%%%%%%%%%%%%%%%%%%%%

In this section we show that spectral relaxation can be interpreted
as a model-based statistical procedure. In particular, we present
a connection between spectral relaxation and Gaussian intrinsic
autoregression models.

Our focus is the spectral relaxation problem presented in
Section~\ref{secpcutrelax}, specifically the constrained eigenvalue
problem in (\ref{eqrelax1}).

Recall that the Laplacian matrix $\mathbf{L}$ is a positive
semidefinite matrix;
moreover, the pseudoinverse $\mathbf{L}^{+}$ is positive semidefinite
and can
be viewed as a kernel matrix. We found this perspective useful in our
discussion of minimum-variance clustering in Section~\ref{secpcutmiv};
note also that~(\cite{SaerensECML2004}) have explored connections
between spectral embedding and random walks on graphs using the
fact that the elements of $\mathbf{L}^{+}$ are closely related to the
commute-time
distances obtained from a random walk on the graph. In this section,
we take the interpretation of $\mathbf{L}^{+}$ in a different direction,
using it to make the connection to Gaussian intrinsic autoregressions.

Denote $\mathbf{K}= \mathbf{L}^{+}$ where $\mathbf{L}=\mathbf{D}-
\mathbf{W}$.
Let us model the $n\times (c - 1)$ matrix $\mathbf{Y}$ as
a singular matrix-variate normal distribution $N_{n, c{-}1}(\mathbf{0},
\sigma^2 \mathbf{K}\otimes\mathbf{I}_{c - 1})$ where we follow
the notation for
matrix-variate normal distributions in (\cite{GuptaNBook2000}).
That is,
\[
p(\mathbf{Y}) \propto\exp\biggl({-}\frac{1}{2\sigma^2}
\operatorname{tr}(\mathbf{Y}'\mathbf{L}\mathbf{Y})\biggr).
\]
Let us set $\sigma^2=1/\operatorname{tr}{(\bolds{\Pi}\mathbf{K})}$
so that
$\mathsf{E}(\mathbf{Y}' \bolds{\Pi}\mathbf{Y}) =
\sigma^2\break\operatorname{tr}(\bolds{\Pi}\mathbf{K}) \mathbf{I}_{c - 1} =
\mathbf{I}
_{c - 1}$.
Finally, we impose the constraint $\mathbf{Y}'\bolds{\Pi}\mathbf{1}_n=\mathbf{0}$
in order to remove the redundancy\break
$\mathbf{K}^{+} \mathbf{1}_n = \mathbf{0}$ in $\mathbf{K}^{+}$.
We thus obtain the following proposition.

\begin{proposition}
The relaxation problem in (\ref{eqrelax1}) is equivalent to the
maximization of the log likelihood $p(\mathbf{Y})$ under the constraints
$\mathbf{Y}' \bolds{\Pi}\mathbf{Y}= \mathbf{I}_{c - 1}$ and
$\mathbf{Y}'\bolds{\Pi}\mathbf{1}_n = \mathbf{0}$.
\end{proposition}

We obtain a statistical interpretation of spectral relaxation
from the fact that a multivariate normal distribution can be
equivalently expressed as a Gaussian conditional autoregression
(CAR) (Besag \citeyear{Besag1974}; Mardia \citeyear{MardiaJMA1988}). Indeed, given
$\mathbf{Y}\thicksim N_{n, c{-}1}\*(\mathbf{0},
\sigma^2 \mathbf{K}\otimes\mathbf{I}_{c - 1})$, we have that
the $\mathbf{y}_i$ can be characterized as $(c - 1)$-dimensional CARs with
%
%e6.1 ###
\begin{eqnarray} \label{eqmcary}
\hspace*{40pt}\mathsf{E}(\mathbf{y}_i|\mathbf{y}_{j}, j\neq i)
&=&\nonumber - \sum_{j\neq i} \frac{l_{ij}}{l_{ii}} \mathbf{y}_j
= \sum_{j= 1}^n \frac{ w_{ij}}{l_{ii} } \mathbf{y}_j, \\[-8pt]\\[-8pt]
\mathsf{Var}(\mathbf{y}_i|\mathbf{y}_{j}, j \neq i)
&=&\nonumber \frac{\sigma^2}{l_{ii}}
\mathbf{I}_{c - 1}. \nonumber
\end{eqnarray}
That is, we have $\mathbf{y}_i|\{\mathbf{y}_{j}\dvtx j\neq i \} \thicksim
N_{c - 1}(\sum_{j= 1}^n\frac{w_{ij}}{ l_{ii}} \mathbf{y}_j,
\break\frac{\sigma^2}{l_{ii}} \mathbf{I}_{c - 1})$, for $i=1, \ldots, n$.
Since $\mathbf{K}$ is positive semidefinite but not positive definite,
\citet{BesagBio1995} referred to such conditional autoregressions
as \textit{Gaussian intrinsic autoregressions}.

The CAR model implicitly requires $w_{ii}=0$ and $l_{ii} =\sum_{j=1}^n
w_{ij}$. In spectral embedding
and clustering~(\cite*{GuatterySIAM2000}; \cite*{Belkin2002}; \cite*{Ng2002}), the
$w_{ij}$ are usually used to assert adjacency or similarity
relationships between the $\mathbf{y}_i$. We will see shortly that these
adjacency or similarity relationships have an interpretation as
conditional independencies.

Since $\mathbf{D}- \mathbf{W}$ is positive semidefinite, $\mathbf{D}- \omega\mathbf{W}$ is positive
definite for $\omega\in(0, 1)$. This fact has been used to devise
CAR models based on $\mathbf{D}- \omega\mathbf{W}$ such that
$\mathsf{E}(\mathbf{y}_i|\mathbf{y}_{j}, j\neq i) = \omega
\sum_{j= 1}^n \frac{ w_{ij}}{l_{ii} }
\mathbf{y}_j$ (see, e.g., \cite*{CarlinBS2003}). We now have
\[
\mathsf{E}(\mathbf{y}_i \mathbf{y}_j'|\mathbf{y}_{l}, l\neq i, j) =
\frac{\omega l_{ij}}{\omega^2 l_{ij}^2 - l_{ii} l_{jj}} \sigma^2
\mathbf{I}_{c - 1}.
\]
As a result, $l_{ij}=0$ (or $w_{ij}=0$) implies that
$\mathbf{y}_i\perp\!\!\!\!\perp\mathbf{y}_j|\{\mathbf{y}_l:l\neq i, j \}$;
that is, $\mathbf{y}_i$ is conditionally independent of $\mathbf{y}_j$
given the remaining vectors. This Markov property also holds for
Gaussian intrinsic autoregressions (\cite*{BesagBio1995}).

This perspective sheds light on some of the relationships between
the \textsc{Ncut} and \textsc{Rcut} formulations of spectral relaxation.
Recall that since $\bolds{\Pi}=\mathbf{D}$ in the \textsc{Ncut}
setting, we impose
the constraints $\mathbf{Y}' \mathbf{D}\mathbf{Y}=\mathbf{I}_{c - 1}$
and $\mathbf{Y}'\mathbf{D}\mathbf{1}_n=\mathbf{0}$. On the
other hand, the \textsc{Rcut} formulation uses the constraints
$\mathbf{Y}' \mathbf{Y}=\mathbf{I}_{c - 1}$ and
\mbox{$\mathbf{Y}'\mathbf{1}_n=\mathbf{0}$} because $\bolds{\Pi}=\mathbf{I}_n$.
Theorem~\ref{THMCEP0}\break shows that the solution of the \textsc{Ncut} is
based on
$\bolds{\Pi}^{-1/2} \mathbf{L}\bolds{\Pi}^{-1/2}
= \mathbf{I}_n - \mathbf{D}^{-1/2} \mathbf{W}\mathbf{D}^{-1/2}$,
which is a so-called
normalized graph Laplacian. The solution of the \textsc{Rcut} problem
is based on the unnormalized graph Laplacian $\mathbf{L}$.
Now
Proposition~\ref{PRO1} reveals a problematic aspect of the \textsc{Ncut}
formulation---piece\-wise constancy of the columns of~$\mathbf{Y}$ is
accompanied by
a lack of orthogonality of these columns. Two natural
desiderata of spectral clustering are in conflict in the \textsc{Ncut}
formulation. This conflict between orthogonality and
piecewise constancy is not present for \textsc{Rcut}. However, the existing
empirical results showed that the normalized graph Laplacian
tends to outperform the unnormalized graph Laplacian. Moreover,
\citet{LuxburgTR2004} provided theoretical evidence of the
superiority of the normalized graph Laplacian.

This seeming paradox can be resolved by using an alternative
choice for $\mathbf{L}$ in the \textsc{Rcut} formulation. Let us set
$\mathbf{L}= (\mathbf{I}_n - \mathbf{C})'(\mathbf{I}_n - \mathbf
{C})$, where $\mathbf{C}=[c_{ij}]$ is
an $n\times n$ nonnegative matrix such that $c_{ii}=0$
for all $i$ and $\mathbf{C}\mathbf{1}_n = \mathbf{1}_n$. Such a
$\mathbf{L}$ is positive
semidefinite but no longer Laplacian. Since $\mathbf{L}\mathbf{1}_n
=\mathbf{0}$,
we can still solve the spectral relaxation problem (\ref{eqrelax2})
using Theorem~\ref{THMCEP0}.

Our experimental results in Section~\ref{secexp2} show that this
novel \textsc{Rcut} formulation is very effective. It is also worth
noting that we can connect this formulation to the simultaneous
autoregression (SAR) model of~\citet{Besag1974}. In particular,
the $\mathbf{y}_i$ are now specified by $n$ simultaneous equations:
\[
\mathbf{y}_i =\sum_{j=1}^n c_{ij} \mathbf{y}_j + \bolds{\varepsilon}_i,\quad
i=1, \ldots, n,
\]
where the $\bolds{\varepsilon}_i$ are independent normal vectors from
$N_{c - 1}(\mathbf{0},
  \sigma^2 \mathbf{I}_{c - 1})$. This equation can be written in
matrix form
as follows:
\begin{eqnarray}
\nonumber\mathbf{Y}= \mathbf{C}\mathbf{Y}+ \bolds{\Sigma}\quad\\
\eqntext{\mbox{with}\quad\bolds{\Sigma}=[\bolds{\varepsilon}_1, \ldots,
\bolds{\varepsilon}_n]' \thicksim N_{n, c{-}1}
\bigl(\mathbf{0},   \sigma^2 \mathbf{I}_n \otimes
\mathbf{I}_{c - 1}\bigr).}
\end{eqnarray}
We thus have
$\mathbf{Y}\thicksim N_{n, c{-}1}
\bigl(\mathbf{0}, \sigma^2 \mathbf{K}\otimes
\mathbf{I}_{c - 1}\bigr)$ with $\mathbf{K}^{+}
= (\mathbf{I}_n - \mathbf{C})'(\mathbf{I}_n - \mathbf{C})$. In practice,
we are especially concerned with the case in which
$\mathbf{C}=\mathbf{D}^{-1}\mathbf{W}$.
It is worth noting that
$\mathbf{I}_n - \mathbf{D}^{-1/2}\mathbf{W}\mathbf{D}^{-1/2}$
and $\mathbf{I}_n-\mathbf{D}^{-1}\mathbf{W}$ have the same
eigenvalues, while the squared
singular values of $\mathbf{I}_n-\mathbf{D}^{-1}\mathbf{W}$ are the
eigenvalues of
$(\mathbf{I}_n-\mathbf{D}^{-1}\mathbf{W})' (\mathbf{I}_n-\mathbf
{D}^{-1}\mathbf{W})$. We thus obtain an interesting
new relationship between the \textsc{Ncut} formulation and the \textsc{Rcut}
formulation.

%%%%%%%%%%%%%%%%%%%%%%%%%%%%%%%%%%%%%%%%%%%%%%%%%%%%%%%%%%%%%%%%%%%%%%%%%%%%%%%
%%%%%%%%%%%%%%%%%%%%%%%%%%%%%%%%%%%%%%%%%%%%%%%%%%%%%%%%%%%%%%%%%%%%%%%%%%%%%%%
%s7 ###
\section{Experiments} \label{secexp2}

Although our principal focus has been to provide a unifying
perspective on spectral clustering, our analysis has
also provided novel spectral algorithms, and it is of interest
to compare the performance of these algorithms to existing algorithms.
In this section we report the results of experiments conducted
with six publicly available data sets: five data sets from the UCI
machine learning repository (the \textit{dermatology} data, the
\textit{vowel} data, the \textit{NIST} optical handwritten digit
data, the \textit{letter} data and the \textit{image segmentation}
data) as well as a set of \textit{gene expression} data analyzed
by~\citet{YeungBio2001}.

In the dermatology data, there are 366 patients, 8 of whom are
excluded due to missing information, with 34 features. The data are
clustered into six classes. We standardized the data to have zero mean
and unit variance. The NIST data set contains the handwritten digits
0--9, where each instance consists of a $16\times 16$~pixel and
where digits are treated as classes. We selected 1000 digits, with
100 instances per digit, for our experiments. The vowel data set
contains the eleven \mbox{steady-state} vowels of British English. The
letter data set consists of images of the letters ``A'' to ``Z.'' In
our experiments we selected the first 10 letters with 195, 199, 182,
207, 203, 210, 226, 196, 188 and 172 instances, respectively. The
image segmentation data consist of seven types of images:
``brickface,'' ``sky,'' ``foliage,'' ``cement,'' ``window,'' ``path'' and
``grass.'' The gene data set contains 384 genes with 17 time points
over two cell cycles. The data were standardized to have mean zero
and unit variance~(\cite*{YeungBio2001}). We treated the five phases
of the cell cycle as five nominal classes for these data,
classifying genes into these classes according to their expression
level peaks. Table~\ref{tabdata} gives a summary of these
data sets.

We compared our rounding algorithm based on Procrustean transformation
(see Algorithm~\ref{alg1}) with those based on the rounding procedures
given in \citet{Bach2006} and \citet{YuICCV2003}, conducting comparisons
using the \textsc{Ncut}, \textsc{Rcut} and minimum-variance criteria.
We refer to the weighted $K$-means and the $K$-means algorithms
of~\citet{Bach2006} as \textit{BJ-wkm} and \textit{BJ-km}, respectively.
Note that the spectral clustering algorithm based on the \textsc{Ncut}
formulation and $K$-means rounding is equivalent to that presented
by~\citet{Ng2002}. We initialized the $K$-means algorithms by the
orthogonal initialization method in \citet{Ng2002}. For the rounding
scheme of~\citet{YuICCV2003}, we used two initialization methods: the
orthogonal initialization method and initialization to the identity
matrix. We refer to the corresponding algorithms as \textit{YS-1} and
\textit{YS-2}. We also used these two initialization methods in
our algorithm (Algorithm~\ref{alg1}), referring to the results
in these two cases as \textit{Margin-1} and \textit{Margin-2}.

%t1 ###
\begin{table}[t]
\tabcolsep=0pt
\caption{\label{tabdata} Summary of the benchmark data sets}
\begin{tabular*}{\columnwidth}{@{\extracolsep{\fill}}@{}ld{3.0}d{3.0}d{3.0}d{4.0}d{4.0}d{4.0}@{}}
\hline
& \multicolumn{1}{@{}c}{\textbf{Gene}} & \multicolumn{1}{c}{\textbf{Dermatology}}
& \multicolumn{1}{c}{\textbf{Vowel}} & \multicolumn{1}{c}{\textbf{NIST}}
& \multicolumn{1}{c}{\textbf{Letter}} & \multicolumn{1}{c@{}}{\textbf{Segmentation}}
\\
\hline
$n$ & $384$ & $358$ & $990$ & $1000$ & $1978$ & $2100$ \\
$d$ & $17$ & $34$ & $10$ & $256$ & $16$ & $19$ \\
$c$ & $5$ & $6$ & $11$ & $10$ & $10$ & $7$ \\
\hline
\end{tabular*}
\tabnotetext[]{tz}{$n$---the number of samples; $d$---the number of features; $c$---the number of classes.}
\end{table}

%%%%%%%%%%%%%%%%%%%%%%%%%%%%%%%%%%%%%%%%%%%%%%%%%%%%%%%%%%%%%%%%%%%
%s7.1 ###
\subsection{Setup and Evaluation Criterion}

We defined the adjacency matrix
$\mathbf{W}=[w_{ij}]$ as $w_{ij}=\exp({-} \|\mathbf{x}_{i} - \mathbf{x}_{j} \|^2/\beta)$
with $\beta>0$.
The kernel matrix is defined as
$\mathbf{K}= \mathbf{H}_n\mathbf{W}\mathbf{H}_n$. For the margin-based
algorithms, however, we set
$w_{ii}=0$ for $i=1, \ldots, n$; in this
case the kernel matrix is defined as
$\mathbf{K}= \mathbf{H}_n(\mathbf{I}_n + \mathbf{W}) \mathbf{H}_n$.
For simplicity, we do not distinguish between these two cases in
our notation in the remainder of this section. In the
minimum-variance formulation we always set
$\bolds{\Pi}=\mathbf{I}_n$. With these
settings, the \textit{BJ-wkm} and \textit{BJ-km} algorithms are based on
the spectral
decomposition of
$\mathbf{I}_n - \mathbf{D}^{-1/2} \mathbf{W}\mathbf{D}^{-1/2}$.
The \textit{YS-1} and \mbox{\textit{YS-2}} algorithms are based on the spectral
decomposition
of $\mathbf{I}_n - \mathbf{D}^{-1} \mathbf{W}$, and the \textit{Margin-1} and \textit{Margin-2} algorithms
are based on the spectral decomposition of
$\mathbf{I}_n - \mathbf{D}^{-1/2} \mathbf{W}\mathbf{D}^{-1/2}$.

Although $\mathbf{L}=\mathbf{D}- \mathbf{W}$ is one natural choice
in the \textsc{Rcut}
setting, we
instead adopted the suggestion in Section~\ref{secgias} and
defined $\mathbf{L}$ as
%
%e7.1 ###
\begin{equation} \label{eqnrcutL}
\mathbf{L}= (\mathbf{I}_n - \mathbf{D}^{-1} \mathbf{W})'(\mathbf{I}_n - \mathbf{D}^{-1} \mathbf{W}).
\end{equation}

To simplify the comparison among procedures, we fixed $\beta$ to
specific sets of values for each of the data sets, exploring a range
of values to investigate the relative sensitivities to the choice
of $\beta$ for the different clustering algorithms. Our specific
choices for both the \textsc{Ncut} and \textsc{Rcut} criteria were
$\beta\in\{1, 10\}$ for the gene data, $\beta\in\{1, 10, 100\}$
for the ``vowel'' data, $\beta\in\{5000, 10000, 20000\}$ for the
``image segmentation'' data, and $\beta\in\{10, 100, 1000\}$ for
the ``dermatology,'' ``NIST'' and ``letter'' data sets. Since the
minimum-variance criterion directly operates on $\mathbf{K}$,
we choose a different set of values when working with
this criterion; in particular, we used $\beta\in\{10, 100\}$
for the gene data, $\beta\in\{100, 1000\}$ for the ``dermatology'' data,
$\beta\in\{1, 10, 100\}$ for the ``vowel'' data, $\beta\in\{500,
1000\}$ for
NIST data, $\beta\in\{10, 100, 1000\}$ for the ``letter'' data, and
$\beta\in\{10, 100, 1000\}$ for the ``image segmentation'' data.

To evaluate the performance of the various clustering algorithms
we employed the Rand index (RI) (\cite*{RandRI1971}). Given a set
of $n$ objects $S = \{O_1, \ldots, O_n\}$, suppose that
$U = \{U_1, \ldots, U_r\}$ and $V = \{V_1, \ldots, V_s\}$ are
two different partitions of the objects in $S$ such that
$\bigcup_{i=1}^r U_i = S = \bigcup_{j=1}^s V_j$ and $U_i
\cap U_{i'} = \varnothing= V_j \cap V_{j'}$ for $i\neq i'$ and $j
\neq j'$. Let $a$ be the number of pairs of objects that are in the
same set in $U$ and in the same set in $V$, and $b$ the number of
pairs of objects that are in different sets in $U$ and in different
sets in $V$. The Rand index is given by $\mathrm{RI} = (a+b)/{n
\choose2 }$. If $\mathrm{RI}=1$, the two partitions are
identical.

Since the ground-truth partitions are available for our six data sets,
we directly calculated RI between the true partition and the
partition obtained from each clustering algorithm. We conducted
50 replicates of each of the algorithms that require random
initialization (this is not necessary for \textit{YS-2} and
\textit{Margin-2}, which are initialized to the identity matrix).
Note that for the \textsc{Rcut} and minimum-variance criteria,
\textit{BJ-wkm} and \textit{BJ-km} become identical because in
these cases $\bolds{\Pi}=\mathbf{I}_n$.

%%%%%%%%%%%%%%%%%%%%%%%%%%%%%%%%%%%%%%%%%%%%%%%%%%%%%%%%%%%%
%%%%%%%%%%%%%%%%%%%%%%%%%%%%%%%%%%%%%%%%%%%%%%%%%%%%%%%%%%%%
%s7.2 ###
\subsection{Performance Analysis}

\begin{figure*}

\includegraphics{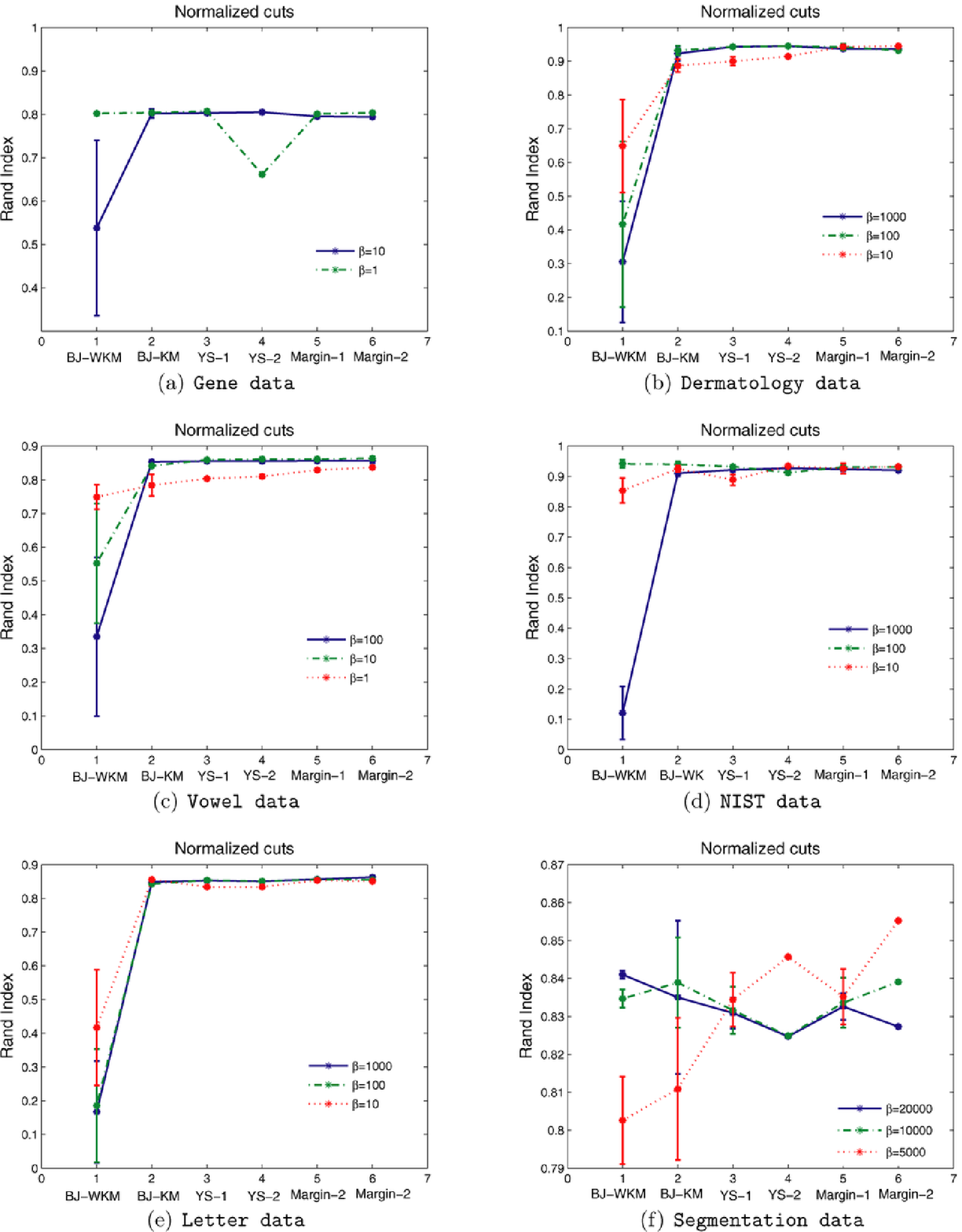}

  \caption{\label{figri1} Clustering results (Rand index) with normalized cuts.
``BJ-WKM'': the weighted $K$-means rounding of Bach and Jordan \protect\citeyear{Bach2006};
``BJ-KM'': the $K$-means rounding of Bach and Jordan \protect\citeyear{Bach2006};
``YS-1'': the rounding scheme of Yu and Shi \protect\citeyear{YuICCV2003}
with the orthogonal initialization method; ``YS-2'': the rounding
scheme of Yu and Shi \protect\citeyear{YuICCV2003} with initialization to the identity
matrix; ``Margin-1'': the rounding scheme in Section~\protect\ref{secprocr}
with the orthogonal initialization method; ``Margin-2'': the rounding
scheme in Section~\protect\ref{secprocr} with initialization to the identity matrix.}
\end{figure*}

\begin{figure*}

\includegraphics{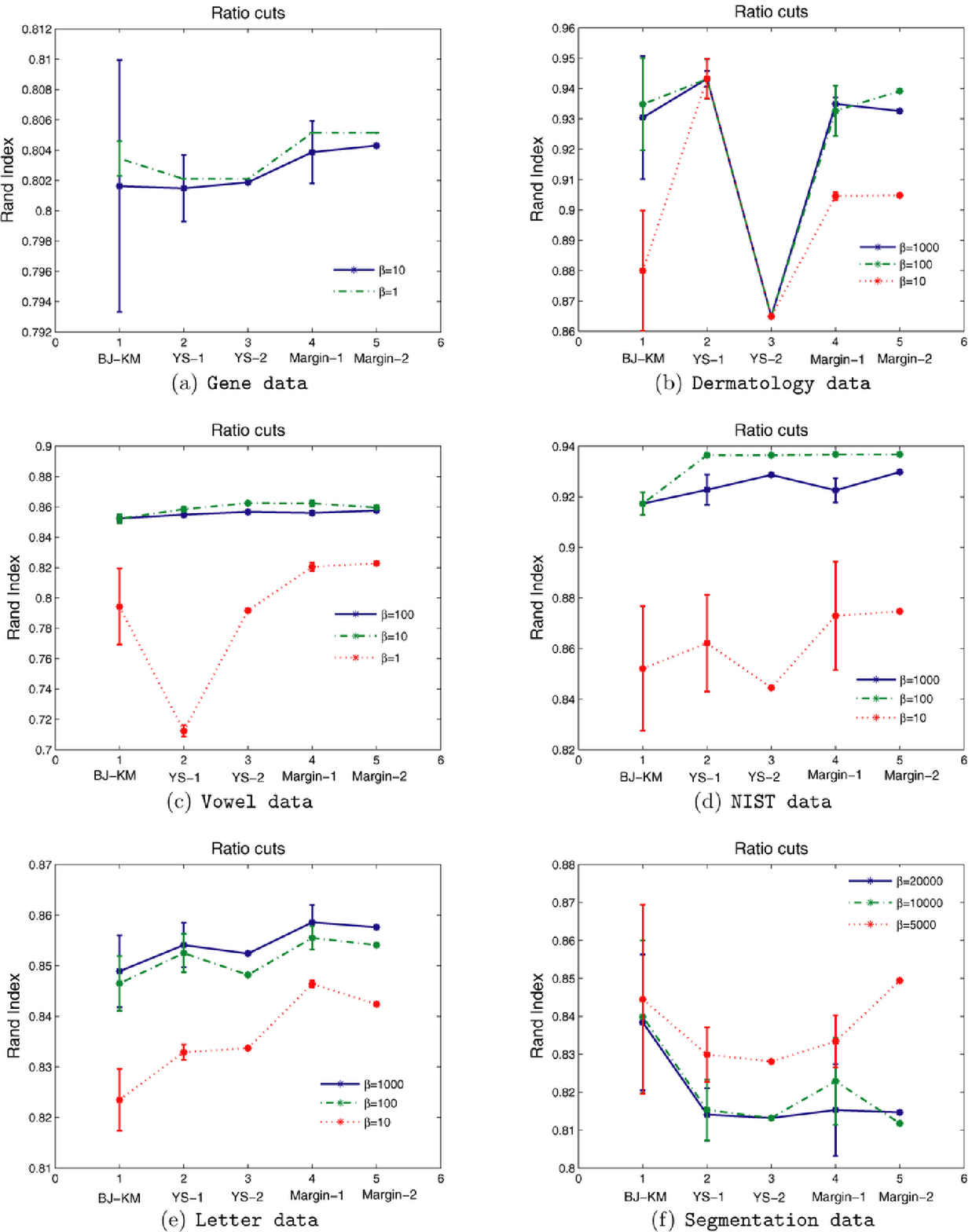}

  \caption{\label{figri2} Clustering results (Rand index) with ratio cuts.
See the caption of Figure~\protect\ref{figri1} for explanation of the acronyms.}
\end{figure*}

\begin{figure*}

\includegraphics{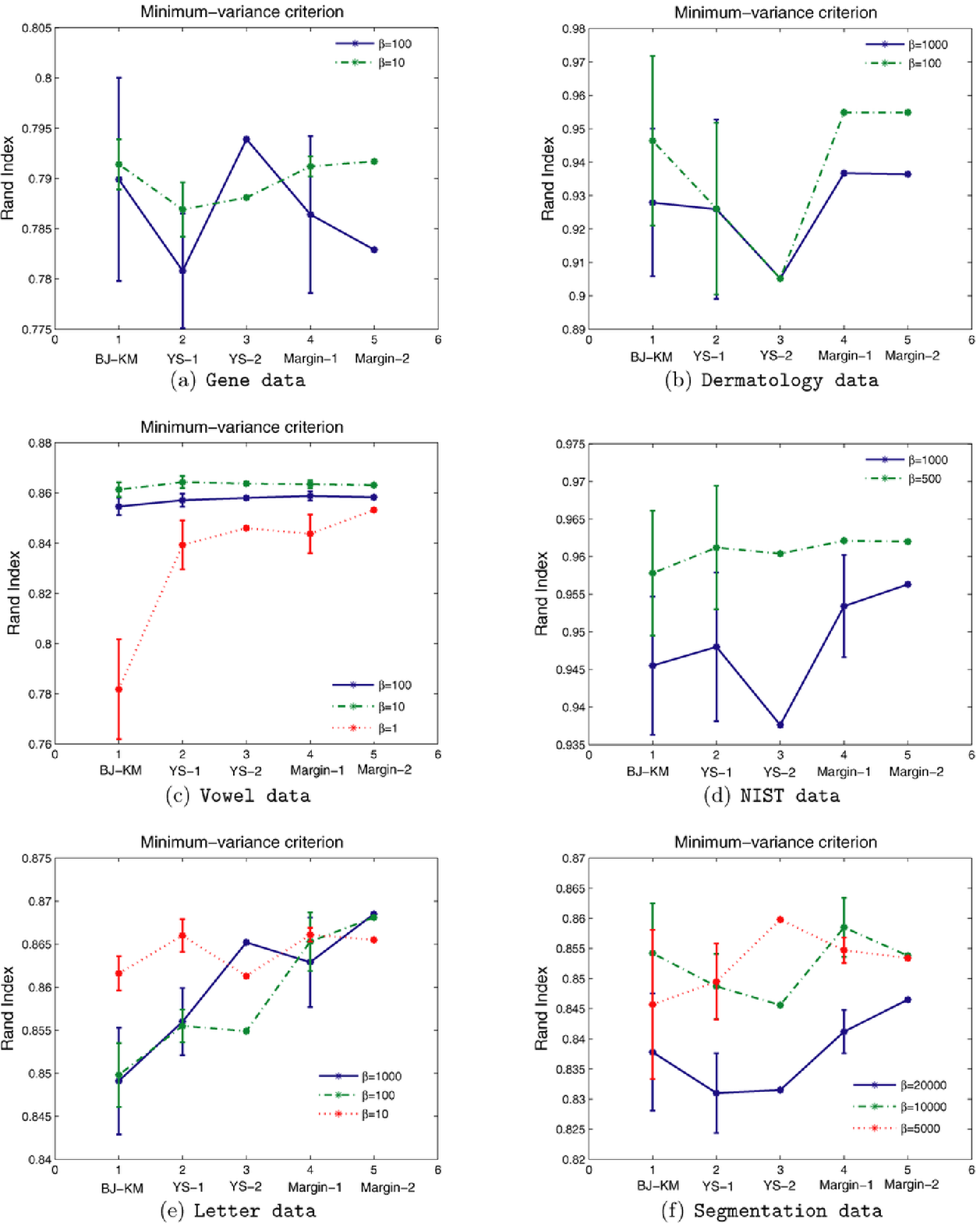}

  \caption{\label{figri3} Clustering results (Rand index)  with the minimum-variance
criterion.  See the caption of Figure~\protect\ref{figri1} for explanation
of the acronyms.}
\end{figure*}

Figure~\ref{figri1} displays the results for all six algorithms
using the \textsc{Ncut} criterion. We see that the margin-based algorithms
are competitive with the other algorithms. The poorest performer in
this setting is \textit{BJ-wkm}, which is highly sensitive to the
value of $\beta$. In particular, when $\beta=10$ for the ``gene''
data set, $\beta\in\{10, 100\}$ for the vowel data,
$\beta\in\{1000, 100, 10\}$
for the ``letter'' data, and $\beta=1000$ for both the ``dermatology''
and ``NIST'' data sets, this algorithm almost failed. A possible
interpretation for this result is the conflict between orthogonality
and piecewise constancy implied in the \textsc{Ncut} setting (see
Proposition~\ref{PRO1}). Indeed, as can be seen from
Figure~\ref{figri1}, the situation is more favorable for the
\textit{BJ-km} rounding algorithm; in this case
$\mathbf{D}^{-12}\mathbf{Y}
(\mathbf{Y}'\mathbf{D}^{-1}\mathbf{Y})^{-1/2}$ is used,
which diminishes the conflict
between orthogonality and piecewise constancy. Similarly, the conflict
is diminished for the \textit{YS} rounding algorithms and our
margin-based rounding methods (because $\mathop{\arg\max}_{j}
d_j^{-1/2} y_{ij}$ is equivalent to $\mathop{\arg\max}_{j} y_{ij}$).

Recall that the \textit{YS-1} and \textit{YS-2} algorithms need to use a
heuristic post-processing procedure; that is, the algorithms operate on
$\hat{\mathbf{Z}}=
\operatorname{dg}({\mathbf{Z}} {\mathbf{Z}}')^{-1/2} {\mathbf{Z}}$. We found that the
performance of the algorithms depends strongly on this procedure.

Figures \ref{figri2} and \ref{figri3} display the experimental
results using the \textsc{Rcut} and minimum-variance criteria, respectively.
We see again that the margin-based algorithms are competitive
with the other algorithms; indeed for several of the data sets
the margin-based algorithms yield better performance than the
other algorithms.

We see from Figures \ref{figri2} and \ref{figri3} that \textit{BJ-km}
is competitive with the other algorithms. This shows that the choice
of $\mathbf{L}$ given in (\ref{eqnrcutL}) is an effective choice.

We again found it to be the case that the heuristic post-processing
procedure was needed for \textit{YS-1} and \textit{YS-2} to yield good
clustering performance.

The performances of \textit{Margin-1} and \textit{Margin-2} were similar
across the data sets and criteria, showing the relative insensitivity
of the margin-based approach to the initialization. Note in particular
the larger degree of variability between the performances of \textit{YS-1}
and \textit{YS-2}. Note also that the margin-based approach was in general
less sensitive to the value of $\beta$ than the other algorithms.

Finally, recall that $\mathbf{L}$ in (\ref{eqnrcutL}) for the \textsc{Rcut}
setting and $\mathbf{L}
=\mathbf{K}^{+}$ obtained from the minimum-variance setting are
positive semidefinite
but they are not Laplacian matrices, because the off-diagonal elements
of the
$\mathbf{W}=\mathbf{L}- \mathbf{D}$
are possibly negative.
Nonetheless, our experimental
results showed that these two choices are still effective. Thus cuts
can be defined through non-Laplacian matrices. Although such cuts lose
their original interpretation in terms of the graph partition, as we have
shown they do have a clear statistical interpretation in terms of Gaussian
intrinsic autoregression models.

%%%%%%%%%%%%%%%%%%%%%%%%%%%%%%%%%%%%%%%%%%%%%%%%%%%%%%%%%%%%%%%%%%%%%%%%%%%
%s8 ###
\section{Discussion} \label{secconclusion}

In this paper we have presented a margin-based perspective on multiway
spectral clustering. We have shown that both aspects of spectral
clustering---\mbox{relaxa-} tion and rounding---can be given an interpretation
in terms of margins. The major advantage of this perspective is that
it ties spectral clustering to the large literature on margin-based
classification. The margin-based perspective has several additional
consequences: (1) it permits a deeper understanding of the relationship
between the normalized cut and ratio cut formulations of spectral
clustering; (2) it strengthens the connections between the minimum-variance
criterion and spectral clustering; and (3) it yields a statistical
interpretation of spectral clustering in terms of Gaussian intrinsic
autoregressions. Also, the preliminary empirical evidence that we
presented suggests that the algorithms motivated by the margin-based
perspective are competitive with existing spectral clustering algorithms.

One of the most useful consequences of the margin-based perspective
is the interpretation that it yields of spectral clustering in
terms of projection onto hyperplanes in a reproducing kernel Hilbert
space (see Figure~\ref{fighyperplane}). This interpretation shows
that the performance of the margin-based clustering algorithms depends
on the separability of the feature vectors. This suggests that the
algorithmic problem of choosing the similarity matrix $\mathbf{W}$ or kernel
matrix $\mathbf{K}$ so as to increase separability is an important
topic for
further research; see~\citet{Bach2006} and \citet{MeilaShi2000} for
initial work along these lines.

Although we have focused on undirected graphs in our treatment,
it is also worth noting the possibility of considering clustering
in a directed graph with the asymmetric weighted matrix
$\mathbf{D}^{-1} \mathbf{W}$ (\cite*{MeilaSAM2007}). This can be
related to
our discussion in Section~\ref{secgias}, where we suggested
the use of the matrix $\mathbf{L}=(\mathbf{I}_n - \mathbf{D}^{-1}
\mathbf{W})'(\mathbf{I}_n - \mathbf{D}^{-1} \mathbf{W})$
in the \textsc{Rcut} setting. The experimental results in
Section~\ref{secexp2} showed that such a suggestion is
promising. Moreover, although $\mathbf{L}$ is no longer Laplacian,
the corresponding spectral relaxation can be interpreted as
a simultaneous autoregression model. The relationship between
simultaneous autoregression and conditional autoregression (\cite*{Ripley1981})
may provide connections between spectral clustering in undirected
graphs and directed graphs. We intend to explore this issue in
future work.

In delineating a relationship between the \textsc{Pcut} criterion and the
kernel minimum-variance criterion, we have proven that the
relaxation problems (\ref{eqrelax1}) and (\ref{eqrelax5}) have the
same solution whenever $\operatorname{rk}(\mathbf{L})=n - 1$ and
$\mathbf{L}^{+}=\mathbf{K}$. This leads
to the question as to whether the original unrelaxed
problems---that is, the minimization of \textsc{Pcut} and the maximization of
$T$ with respect to discrete partition matrix $\mathbf{E}$--- have the same
solution under the conditions $\operatorname{rk}(L)=n - 1$ and
$\mathbf{L}^{+}=\mathbf{K}$. This
is currently an open problem.

%%%%%%%%%%%%%%%%%%%%%%%%%%%%%%%%%%%%%%%%%%%%%%%%%%%%%%%%%%%%%%%%%%%%%%%%
%%%%%%%%%%%%%%%%%%%%%%%%%%%%%%%%%%%%%%%%%%%%%%%%%%%%%%%%%%%%%%%%%%%%%%%%
\begin{appendix}

%s9 ###
\section*{Appendix}\label{ap:relaxation}
\subsection{\texorpdfstring{Proof of Proposition~\protect\ref{PRO1}}{Proof of Proposition~1}}\label{aprelaxation}

Since the columns of $\mathbf{Y}$ are piecewise constant
with respect to the partition $\mathbf{E}$,
we can express
$\mathbf{Y}$ as $\mathbf{Y}= \mathbf{E}\bolds{\Psi}$
for some
$\bolds{\Psi}\in\mathbb{R}^{c\times {(c - 1)}}$.
Let
$\mathbf{Y}_0 = \bolds{\Pi}^{1/2} \mathbf{Y}$,
$\bolds{\Psi}_0 = [\bolds{\Psi}, \alpha\mathbf{1}_c]$, a
$c\times c$
matrix, and
$\mathbf{Z}=[\mathbf{Y}_0, \alpha\bolds{\Pi}^{1/2}\mathbf{1}_n]$,
where
$\alpha=1/\sqrt{\mathbf{1}_n' \bolds{\Pi}\mathbf{1}_n}$.
We have
$\bolds{\Pi}^{-1/2}\mathbf{Z}
= \mathbf{E}\bolds{\Psi}_0$ and $\mathbf{Z}' \mathbf{Z}= [\mathbf{Y}_0, \alpha\bolds{\Pi}^{1/2}
\mathbf{1}_n]' [\mathbf{Y}_0, \alpha\bolds{\Pi}^{1/2}
\mathbf{1}_n] = \mathbf{I}_c$ due to $\mathbf{E}\mathbf{1}_c =\mathbf{1}_n$, $\mathbf{Y}_0'
\mathbf{Y}_0 = \mathbf{Y}' \bolds{\Pi}\mathbf{Y}= \mathbf{I}_{c - 1}$
and $\mathbf{Y}_0'
\bolds{\Pi}^{1/2} \mathbf{1}_n = \mathbf{Y}' \bolds{\Pi}\mathbf{1}_n = \mathbf{0}$.
Furthermore, we have
$\bolds{\Psi}_0' \mathbf{E}' \bolds{\Pi}\mathbf{E}\bolds{\Psi}_0
= \mathbf{Z}' \mathbf{Z}= \mathbf{I}_{c}$. Since
$\bolds{\Psi}_0$
and
$\mathbf{E}' \bolds{\Pi}\mathbf{E}$
are square,
$\bolds{\Psi}_0$
and
$\mathbf{E}' \bolds{\Pi}\mathbf{E}$
are invertible. Hence
$\bolds{\Psi}_0 \bolds{\Psi}_0' = (\mathbf{E}' \bolds{\Pi}\mathbf{E})^{-1}$.
We now have
\begin{eqnarray*}
\operatorname{tr}(\mathbf{Y}' \mathbf{L}\mathbf{Y})
&=&
\operatorname{tr}(\mathbf{Y}_0' \bolds{\Pi}^{-1/2}
\mathbf{L}\bolds{\Pi}^{-1/2} \mathbf{Y}_0)
\\
&=& \operatorname{tr}(\mathbf{Z}'\bolds{\Pi}^{-1/2}\mathbf{L}\bolds{\Pi}^{-1/2}\mathbf{Z})
= \operatorname{tr}(\bolds{\Psi}_0' \mathbf{E}'
\mathbf{L}\mathbf{E}\bolds{\Psi}_0)
\\
&=& \operatorname{tr}( \mathbf{E}' \mathbf{L}\mathbf{E}\bolds{\Psi}_0
\bolds{\Psi}_0')
=
\operatorname{tr}( \mathbf{E}' \mathbf{L}\mathbf{E}(\mathbf{E}'
\bolds{\Pi}\mathbf{E})^{-1}),
\end{eqnarray*}
completing the proof. %\qed

%%%%%%%%%%%%%%%%%%%%%%%%%%%%%%%%%%%%%%%%%%%%%%%%%%%%%%%%%%%%%%%%%%%%%%%%
%
%s10 ###
\subsection{\texorpdfstring{The Proof of Proposition~\protect\ref{PROEXIST}}{The Proof of Proposition~2}}\label{apexistence}

In this section we provide a constructive proof of
Proposition~\ref{PROEXIST} by establishing the existence of $\bolds
{\Psi}$.
We also provide an example of the construction in the special
case of $c = 4$ and $\bolds{\Pi}= \mathbf{I}_{n}$.

Let $(\mathbf{E}' \bolds{\Pi}\mathbf{E})^{-1} =
\operatorname{diag}(1/\beta_1, \ldots, 1/\beta_c)$
and $\bolds{\beta}=(\beta_1,\break \ldots, \beta_c)'$. We then have
$\mathbf{1}_n' \bolds{\Pi}\mathbf{1}_n =
\bolds{\pi}'\mathbf{1}_n = \bolds{\beta}' \mathbf{1}
_c$ and $\mathbf{E}' \bolds{\Pi}\mathbf{1}_n = \bolds{\beta}$.
In the proof in Appendix~\ref{aprelaxation}, we obtain
$\bolds{\Psi}_0 \bolds{\Psi}_0' = (\mathbf{E}' \bolds{\Pi}\mathbf{E})^{-1}$.
Thus,
\begin{eqnarray}
\bolds{\Psi}\bolds{\Psi}' = \operatorname{diag}(1/\beta_1, \ldots, 1/\beta_c)
- \frac{1}{\bolds{\pi}'\mathbf{1}_n} \mathbf{1}_c \mathbf{1}_c'\nonumber\\
\eqntext{(\mbox{denoted }\mathbf{A}).}
\end{eqnarray}
In order to make the above equation hold, it is necessary for
$\mathbf{A}$ to be positive semidefinite. Given any nonzero
$\mathbf{b}=(b_1, \ldots, b_c)'\in\mathbb{R}^c$, we have
\begin{eqnarray*}
&&\mathbf{b}' \operatorname{diag}
(\bolds{\beta})\mathbf{A}\operatorname{diag}
(\bolds{\beta})\mathbf{b}/(\bolds{\pi}
'\mathbf{1}_n)
\\
&&\quad = \sum_{j=1}^c \frac{\beta_j}{\bolds{\pi}
'\mathbf{1}_n} b_j^2 -\biggl( \sum_{j=1}
\frac{\beta_j}{\bolds{\pi}'\mathbf{1}_n} b_j \biggr)^2 \geq0,
\end{eqnarray*}
since the function $f(x)=x^2$ is convex. This implies that
$\mathbf{A}$ positive semidefinite. Furthermore, it is easy to obtain
$\mathbf{A}\bolds{\beta}
= \mathbf{0}$.
Using the SVD of $\mathbf{A}$, we are always able to obtain a $\bolds
{\Psi}$ such
that $\bolds{\Psi}\bolds{\Psi}' = \mathbf{A}$ and $\bolds{\Psi}'
\bolds{\beta}= \mathbf{0}$. Consequently, we have
\[
\mathbf{1}_n' \bolds{\Pi}\mathbf{E}\bolds{\Psi}= \bolds{\beta}'
\bolds{\Psi}=0 \quad\mbox{and}\quad\bolds{\Psi}' \mathbf{E}' \bolds{\Pi}\mathbf{E}\bolds{\Psi}
= \mathbf{I}_{c - 1}.
\]
The latter equality comes from
\begin{eqnarray*}
\mathbf{I}_c &=& \bolds{\Psi}_0'\mathbf{E}' \bolds{\Pi}\mathbf{E}\bolds{\Psi}_0
=
\left[
\matrix{
\bolds{\Psi}'
\cr \alpha\mathbf{1}_c'
}
\right]
\mathbf{E}' \bolds{\Pi}\mathbf{E}[\bolds{\Psi}, \alpha\mathbf{1}_c]
\\
&=& \left[
\matrix{
\bolds{\Psi}' \mathbf{E}' \bolds{\Pi} \mathbf{E}\bolds{\Psi}& \mathbf{0}
\cr \mathbf{0} & 1
}
\right].
\end{eqnarray*}

\begin{example} \label{exm1}
Let $\eta=\bolds{\pi}'\mathbf{1}_n$ and
$\eta_j
=\sum_{i \in V_j} \pi_i$.
Assume that $\bolds{\Psi}=(\bolds{\psi}_1, \ldots, \bolds{\psi}_{c - 1})'$ where
$\bolds{\psi}_1'=\break(\frac{\sqrt{\eta-\eta_1}}{\sqrt{\eta\eta_1}},
-\frac{\sqrt{\eta_1}}{\sqrt{\eta(\eta-\eta_1)}} \mathbf{1}_{c - 1}'
)$
and
\begin{eqnarray*}
&&\bolds{\psi}_l'=\biggl(0 * \mathbf{1}_{l{-}1}', \frac{\sqrt{\sum_{j=l+1}^c \eta_j}}
{\sqrt{\eta_{l} \sum_{j=l}^c \eta_j}},\\
&&\hspace*{35pt}\frac{\sqrt{\eta_l}}{\sqrt{ \sum_{j=l}^c \eta_j \sum_{j=l+1}^c
\eta_j}} \mathbf{1}_{c{-}l} \biggr)
\end{eqnarray*}
for $l=2, \ldots, c - 1$.
For instance, if $c=4$, we have
\begin{eqnarray*}
\bolds{\Psi}=
\fontsize{11}{11}\selectfont{
\left[
\matrix{
\tfrac{\sqrt{\eta-\eta_1}}{\sqrt{\eta\eta_1}}& 0 & 0 \cr
- \tfrac{\sqrt{\eta_1}}{\sqrt{\eta(\eta{-}\eta_1)}} & \tfrac
{\sqrt{\eta_3{+}\eta_4}}{\sqrt{\eta_2(\eta{-} \eta_1)}} & 0 \cr
- \tfrac{\sqrt{\eta_1}}{\sqrt{\eta(\eta{-}\eta_1)}} & - \tfrac
{\sqrt{\eta_2}}{\sqrt{(\eta_3{+}\eta_4)(\eta{-} \eta_1)}} &
\tfrac{\sqrt{\eta_4}}{\sqrt{(\eta_3{+}\eta_4)\eta_3}} \cr
- \tfrac{\sqrt{\eta_1}}{\sqrt{\eta(\eta{-}\eta_1)}} & - \tfrac
{\sqrt{\eta_2}}{\sqrt{(\eta_3{+}\eta_4)(\eta{-} \eta_1)}}
&-\tfrac{\sqrt{\eta_3}}{\sqrt{(\eta_3{+}\eta_4)\eta_4}}}%
\right]
}
.
\end{eqnarray*}
It is easily verified that $\mathbf{Y}= \mathbf{E}\bolds{\Psi}$
satisfies the conditions (a)--(c)
listed in Proposition~\ref{PRO1}. Let $\mathbf{a}_1, \ldots , \mathbf{a}_c$
denote the
row vectors of $\bolds{\Psi}$. We note that an arbitrary collection of
$c - 1$ vectors from the set  $\mathbf{a}_1, \ldots , \mathbf{a}_c$ are linearly
independent. The convex hull of  $\mathbf{a}_1, \ldots , \mathbf{a}_c$ is
thus a
$(c - 1)$-dimensional simplex. (A $d$-dimensional simplex is
the convex hull of an affinely independent point set in~$\mathbb{R}^d$. A
regular $d$-dimensional simplex is the convex hull of $d + 1$ points
with all pairs of points having equal distances.) In addition, we
have that the squared distance between $\mathbf{a}_i$ and $\mathbf{a}_j$ is
\[
\|\mathbf{a}_i-\mathbf{a}_j\|^2 = \frac{1}{\eta_i} + \frac{1}{\eta_j}
\quad\mbox{for }
i\neq j.
\]
Note that we have $\eta=n$ and $\eta_j=n_j$ when $\bolds{\Pi
}=\mathbf{I}_n$.
In particular, if $\bolds{\Pi}=\mathbf{I}_n$ and $n_1=\cdots
=n_c=\frac{n}{c}$, the
$\mathbf{a}_i$
constitute the vertices of a $(c - 1)$-dimensional regular simplex.
\end{example}

%s11 ###
\subsection{\texorpdfstring{The Proof of Theorem~\protect\ref{THMCEP0}}{The Proof of Theorem~1}}\label{aplagrangian}

This theorem is a variation on a standard result in linear
algebra; for completeness we present a proof. Let
$\mathbf{S}= \bolds{\Pi}^{-1/2} \mathbf{L}\bolds{\Pi}^{-1/2}$ and
consider the following Lagrangian:
\begin{eqnarray*}
&&\hspace*{-6pt}L(\mathbf{Y}_0, \mathbf{A}, \mathbf{b})
\\
&&\ \ \hspace*{-6pt}=\operatorname{tr}\bigl(\mathbf{Y}_0' \mathbf{S}\mathbf{Y}_0\bigr) -
\operatorname{tr}(\mathbf{A}(\mathbf{Y}_0'\mathbf{Y}_0-\mathbf{I}_{c - 1}))
- \mathbf{b}'\mathbf{Y}_0' \bolds{\Pi}^{1/2} \mathbf{1}_n,
\end{eqnarray*}
where $\mathbf{A}$ is a $(c - 1)\times (c - 1)$ symmetric matrix of
Lagrange multipliers and $\mathbf{b}$ is a $(c - 1)\times 1$ vector of
Lagrange multipliers. We differentiate to obtain
\[
\frac{\partial L} {\partial\mathbf{Y}_0} = 2 \mathbf{S}\mathbf{Y}_0
- 2 \mathbf{Y}_0 \mathbf{A}-\bolds{\Pi}^{1/2} \mathbf{1}_n \mathbf{b}'.
\]
Letting $\frac{\partial L} {\partial\mathbf{Y}_0} = \mathbf{0}$
leads to
\[
2 \mathbf{S}\mathbf{Y}_0 - 2 \mathbf{Y}_0 \mathbf{A}-
\bolds{\Pi}^{1/2} \mathbf{1}_n \mathbf{b}' =\mathbf{0},
\]
from which we have
\[
2 \mathbf{1}_n'\bolds{\Pi}^{1/2} \mathbf{S}\mathbf{Y}_0 -
2 \mathbf{1}_n'\bolds{\Pi}^{1/2} \mathbf{Y}_0
\mathbf{A}- \mathbf{1}_n'\bolds{\Pi}\mathbf{1}_n \mathbf{b}' =\mathbf{0}.
\]
This implies $\mathbf{b}=\mathbf{0}$. Accordingly, we obtain
\[
\mathbf{S}\mathbf{Y}_0 = \mathbf{Y}_0 \mathbf{A}.
\]
We now take the eigendecomposition of $\mathbf{A}$, letting
$\mathbf{A}= \mathbf{Q}' \bolds{\Gamma}_1\mathbf{Q}$ where
$\mathbf{Q}$ is a $(c - 1)\times (c - 1)$
orthonormal matrix and
$\bolds{\Gamma}_1$ is a $(c - 1)\times (c - 1)$ diagonal matrix. We note
that the diagonal entries of $\bolds{\Gamma}_1$ and the columns of
$\mathbf{Y}_0 \mathbf{Q}'$
are the eigenvalues and the associated eigenvectors of $\mathbf{S}$.
Clearly, $\bolds{\Pi}^{1/2} \mathbf{1}_n$ is the
eigenvector of $\mathbf{S}$
associated with eigenvalue $0$. We now let
$\bolds{\Gamma}_1=\operatorname{diag}(\gamma_2,\ldots, \gamma_c)$.
We thus have $\bar{\mathbf{Y}}_0 =[\bolds{\mu}_2, \ldots,\bolds{\mu}_c] \mathbf{Q}$.
Obviously, $\bar{\mathbf{Y}}_0$ satisfies
$\bar{\mathbf{Y}}_0'\bar{\mathbf{Y}}_0 = \mathbf{I}_{c - 1}$ and
\mbox{$\bar{\mathbf{Y}}_0'\bolds{\Pi}^{1/2} \mathbf{1}_n = \mathbf{0}$} due to
$\bolds{\mu}_i' \bolds{\Pi}^{1/2}\mathbf{1}_n = 0$ for $i\neq1$.

To verify that $\bar{\mathbf{Y}}_0$ is the solution of problem
(\ref{eqrelax2}), we consider the Hessian matrix of $L$ with
respect to $\mathbf{Y}_0$. Let $\operatorname{vec}(\mathbf{Y}_0')=(y_{11}, \ldots, y_{1,c{-}1},
y_{21}, \ldots, y_{n, c{-}1})'$. The Hessian matrix is then given by
\[
\mathbf{H}(\mathbf{Y}_0) = \frac{\partial^2 L}
{{\partial\operatorname{vec}(\mathbf{Y}_0')}\,{\partial\operatorname{vec}(\mathbf{Y}_0')}'}
= \mathbf{I}_{c - 1} \otimes \mathbf{S}- \mathbf{A}\otimes  \mathbf{I}_{n}.
\]
Let $\mathbf{B}$ be an arbitrary nonzero $n\times (c - 1)$ matrix
such that
$\mathbf{B}' [\bolds{\mu}_1, \ldots, \bolds{\mu}_c]=\mathbf{0}$.
We can always express $\mathbf{B}=
[\bolds{\mu}_{c{+}1}, \ldots, \bolds{\mu}_n] \bolds{\Phi}$ where
$\bolds{\Phi}=[\bolds{\phi}_1, \ldots,\bolds{\phi}_{c - 1}]$ is an
$(n - c)\times (c - 1)$ matrix. Denoting
$\bolds{\Gamma}_2=\operatorname{diag}(\gamma_{c{+}1}, \ldots,\break\gamma_n)$, we have
\begin{eqnarray*}
&&\operatorname{vec}((\mathbf{B}\mathbf{Q})')' \mathbf{H}
(\bar{\mathbf{Y}}_0) \operatorname{vec}((\mathbf{B}\mathbf{Q})')
\\
&&\ =\operatorname{tr}( \mathbf{Q}' \mathbf{B}' \mathbf{S}\mathbf{B}\mathbf{Q})
- \operatorname{tr}(\mathbf{A}\mathbf{Q}' \mathbf{B}' \mathbf{B}\mathbf{Q})
\\
&&\ = \operatorname{tr}(\mathbf{B}' \mathbf{S}\mathbf{B})
\,{-}\, \operatorname{tr}(\bolds{\Gamma}_1 \mathbf{B}' \mathbf{B})
\,{=}\,
\operatorname{tr}(\bolds{\Phi}' \bolds{\Gamma}_2 \bolds{\Phi})\,{-}\,
\operatorname{tr}(\bolds{\Gamma}_1 \bolds{\Phi}' \bolds{\Phi})
\\
&&\ = \sum_{i=1}^{c - 1} \bolds{\phi}_i' \bolds
{\Gamma}_2 \bolds{\phi}_i - \sum_{i=1}^{c - 1}
\gamma_{i+1} \bolds{\phi}_i'
\bolds{\phi}_i
\\
&&\ = \sum_{i=1}^{c - 1} \bolds{\phi}_i'(\bolds{\Gamma}_2
- \gamma_{i+1} \mathbf{I}_{n{-}c})\bolds{\phi}_i \geq0.
\end{eqnarray*}
If $\gamma_c> \gamma_{c+1}$, then the matrices $\bolds{\Gamma}_2 -
\gamma_{i+1}
\mathbf{I}_{n{-}c}$, $i=1, \ldots, c - 1$, are positive definite.
Thus, the
above inequality is strict. This shows that $\bar{\mathbf{Y}}_0$ is a
strict local minimum of $\operatorname{tr}(\mathbf{Y}_0'\bolds{\Pi}^{-1/2} \mathbf{L}
\bolds{\Pi}^{-1/2} \mathbf{Y}_0)$ under the conditions
$\mathbf{Y}_0'\mathbf{Y}_0=\mathbf{I}_{c - 1}$ and $\mathbf{Y}_0'\bolds{\Pi}^{1/2}
\mathbf{1}_n=\mathbf{0}$.

%s12 ###
\subsection{\texorpdfstring{The Solution of Problem~(\protect\ref{EQRELAX6})}{The Solution of Problem~(4.3)}}\label{aprkhs}

Let $\mathbf{T}= \bolds{\Pi}^{1/2} \mathbf{H}_{\pi}'
\mathbf{K}\mathbf{H}_{\pi}\bolds{\Pi}^{1/2}$ and consider the following Lagrangian:
\begin{eqnarray*}
&&L(\mathbf{Y}_0, \mathbf{A}, \mathbf{b})
\\
&&\quad= \operatorname{tr}(\mathbf{Y}_0' \mathbf{T}\mathbf{Y}_0)
-\operatorname{tr}\bigl(\mathbf{A}(\mathbf{Y}_0'\mathbf{Y}_0-\mathbf{I}_{c - 1})\bigr)
- \mathbf{b}'\mathbf{Y}_0' \bolds{\Pi}^{1/2} \mathbf{1}_n,
\end{eqnarray*}
where $\mathbf{A}$ is a $(c - 1)\times (c - 1)$ symmetric matrix of
Lagrange multipliers and $\mathbf{b}$ is a $(c - 1)\times 1$ vector of
Lagrange multipliers. Differentiating, we obtain
\[
\frac{\partial L} {\partial\mathbf{Y}_0} = 2 \mathbf{T}\mathbf
{Y}_0 - 2 \mathbf{Y}_0 \mathbf{A}-
\bolds{\Pi}^{1/2} \mathbf{1}_n \mathbf{b}'.
\]
Letting $\frac{\partial L} {\partial\mathbf{Y}_0} = \mathbf{0}$
leads to
\[
2 \mathbf{T}\mathbf{Y}_0 - 2 \mathbf{Y}_0 \mathbf{A}- \bolds{\Pi
}^{1/2} \mathbf{1}_n \mathbf{b}' =\mathbf{0},
\]
from which we have
\[
2 \mathbf{1}_n'\bolds{\Pi}^{1/2} \mathbf{T}\mathbf{Y}_0 -
2 \mathbf{1}_n'\bolds{\Pi}^{1/2} \mathbf{Y}_0
\mathbf{A}- 1_n'\bolds{\Pi}\mathbf{1}_n \mathbf{b}' =\mathbf{0}.
\]
Since $\mathbf{1}_n'\bolds{\Pi}^{1/2} \mathbf{T}= \mathbf
{1}_n' \bolds{\Pi}\mathbf{H}_{\pi}' \mathbf{K}\mathbf{H}_{\pi}
\bolds{\Pi}^{1/2} = \mathbf{1}_n' \mathbf{H}_{\pi}
\bolds{\Pi}\mathbf{K}\cdot\break\mathbf{H}_{\pi}
\bolds{\Pi}^{1/2} = \mathbf{0}$, we obtain $\mathbf{b}=\mathbf{0}$.
This implies
\[
\mathbf{T}\mathbf{Y}_0 = \mathbf{Y}_0 \mathbf{A}.
\]
Now following the proof in Appendix~\ref{aplagrangian},
we find that the top $c - 1$ eigenvectors of $\mathbf{T}$ provide
the solution for $\mathbf{Y}_0$ in problem~(\ref{EQRELAX6}).

%s13 ###
\subsection{\texorpdfstring{The Proof of Theorem~\protect\ref{THM6}}{The Proof of Theorem~3}}\label{apmp}

Our proof is based on the following lemma.

\begin{lemma} \label{LEM2} Assume that $\mathbf{A}$ is an $n\times n$
symmetric matrix
with $\operatorname{rk}(\mathbf{A})=n - 1$ and
$\mathbf{A}\mathbf{1}_n =\mathbf{0}$.
Let $\mathbf{A}^{+}$ be the MP
inverse of $\mathbf{A}$. Then
$\bolds{\Pi}^{1/2} \mathbf{H}_{\pi}' {\mathbf{A}^{+}}\mathbf{H}_{\pi} \bolds{\Pi}^{1/2}$
is the MP inverse of
$\bolds{\Pi}^{-1/2}\mathbf{A}\bolds{\Pi}^{-1/2}$.
\end{lemma}

\begin{pf}
We first prove $\mathbf{A}^{+}\mathbf{A}= \mathbf{A}\mathbf{A}^{+}
= \mathbf{H}_n$. Let $\mathbf{N}= \mathbf{A}' \mathbf{A}$. It
is clear that
$\mathbf{N}\mathbf{H}_{n} = \mathbf{H}_{n} \mathbf{N}= \mathbf{N}$. It thus follows from
Corollary 4.5.18 in \citet{Horn1985} that there exists an
$n\times n$ orthonormal matrix $\mathbf{U}$ such that
\[
\mathbf{U}' \mathbf{N}\mathbf{U}\,{=}\,
\pmatrix{
\bolds{\Delta}_{n - 1} & \mathbf{0} \cr
\mathbf{0} & 0
}
\ \ \mbox{and}\ \ \mathbf{U}' \mathbf{H}_{n} \mathbf{U}
\,{=}\,
\pmatrix{
\mathbf{I}_{n - 1} & \mathbf{0} \cr
\mathbf{0} & 0
}
,
\]
where $\bolds{\Delta}_{n - 1}$ is an $(n - 1)\times (n - 1)$
diagonal matrix
with positive diagonal entries, and
$\mathbf{U}=[\mathbf{U}_1, \frac{1}{\sqrt{n}}\mathbf{1}_n]$
with
$\mathbf{U}_1' \mathbf{U}_1 = \mathbf{I}_{n - 1}$
and $\mathbf{U}_1 \mathbf{1}_n =\mathbf{0}$. Here we
use the fact that $\mathbf{1}_n$ is the eigenvector of
$\mathbf{N}$
and of $\mathbf{H}_{n}$
with associated eigenvalue $0$. Accordingly, we have
\[
\mathbf{N}= \mathbf{U}_1 \bolds{\Delta}_{n - 1} \mathbf{U}_1'
\quad\mbox{and}\quad\mathbf{H}_{n} = \mathbf{U}_1 \mathbf{U}_1',
\]
from which it follows that
\[
\mathbf{N}^{+} = \mathbf{U}_1 \bolds{\Delta}_{n - 1}^{-1} \mathbf{U}_1'
\]
and hence
$\mathbf{N}^{+} \mathbf{N}= \mathbf{U}_1 \mathbf{U}_1' =\mathbf{H}_{n}$. On the other hand,
since
$\mathbf{A}^{+} = (\mathbf{A}'\mathbf{A})^{+} \mathbf{A}'$,
we have
$\mathbf{A}^{+} \mathbf{A}= \mathbf{N}^{+} \mathbf{N}=\mathbf{H}_{n}$.
Since $\mathbf{A}$ is symmetric, we also have
$\mathbf{A}\mathbf{A}^{+} = \mathbf{H}_{n}$.

Using the identity
$\mathbf{A}^{+}\mathbf{A}= \mathbf{A}\mathbf{A}^{+} = \mathbf{H}_{n}$
and $\mathbf{A}\mathbf{H}_{\pi}' = \mathbf{A}= \mathbf{H}_{\pi} \mathbf{A}$, we have
\begin{eqnarray*}
&&\bolds{\Pi}^{-1/2} \mathbf{A}\bolds{\Pi}^{-1/2} \bolds{\Pi}^{1/2}
\mathbf{H}_{\pi}' {\mathbf{A}^{+}} \mathbf{H}_{\pi}
\bolds{\Pi}^{1/2}
\\
&&\quad= \bolds{\Pi}^{-1/2} \mathbf{H}_{\pi} \bolds{\Pi}^{1/2}
=\bolds{\Pi}^{1/2} \mathbf{H}_{\pi}' \bolds{\Pi}^{-1/2}
\\
&&\quad= \bolds{\Pi}^{1/2}
\mathbf{H}_{\pi}' \mathbf{A}^{+} \mathbf{H}_{\pi} \bolds{\Pi}^{1/2}
\bolds{\Pi}^{-1/2} \mathbf{A}\bolds{\Pi}^{-1/2}.
\end{eqnarray*}
We further obtain
\begin{eqnarray*}
&&\bolds{\Pi}^{-1/2} \mathbf{A}\bolds{\Pi}^{-1/2}
\bolds{\Pi}^{1/2} \mathbf{H}_{\pi}' \mathbf{A}^{+} \mathbf{H}_{\pi} \bolds{\Pi}^{1/2}
\bolds{\Pi}^{-1/2} \mathbf{A}\bolds{\Pi}^{-1/2}
\\
&&\quad= \bolds{\Pi}^{-1/2}\mathbf{A}\bolds{\Pi}^{-1/2}
\end{eqnarray*}
and
\begin{eqnarray*}
&&\bolds{\Pi}^{1/2} \mathbf{H}_{\pi}' \mathbf{A}^{+}
\mathbf{H}_{\pi} \bolds{\Pi}^{1/2}\bolds{\Pi}^{-1/2} \mathbf{A}\bolds{\Pi}^{-1/2}
\bolds{\Pi}^{1/2} \mathbf{H}_{\pi}' \mathbf{A}^{+}
\\
&&\quad
{}\cdot\mathbf{H}_{\pi}\bolds{\Pi}^{1/2}
= \bolds{\Pi}^{1/2} \mathbf{H}_{\pi}' \mathbf{A}^{+}\mathbf{H}_{\pi} \bolds{\Pi}^{1/2}.
\end{eqnarray*}
Thus $\bolds{\Pi}^{1/2} \mathbf{H}_{\pi}' {\mathbf{A}^{+}} \mathbf{H}_{\pi}
\bolds{\Pi}^{1/2}$ is the MP inverse of $\bolds{\Pi}^{-1/2} \break\cdot\mathbf{A}\bolds{\Pi}^{-1/2}$.
\end{pf}

Since ${\mathbf{L}^{+}}$ is the MP inverse of
$\mathbf{L}$, ${\mathbf{L}^{+}}$ is
positive semidefinite and it satisfies
${\mathbf{L}^{+}} \mathbf{1}_n= \mathbf{0}$ and
$\operatorname{rk}({\mathbf{L}^{+}})=n - 1$. It is obvious that
$\operatorname{rk}\bigl(\bolds{\Pi}^{-1/2} \mathbf{L}\bolds{\Pi}^{-1/2}\bigr) = n - 1$
and
$\operatorname{rk}\bigl(\bolds{\Pi}^{1/2} \mathbf{H}_{\pi}'
{\mathbf{L}^{+}} \mathbf{H}_{\pi}\bolds{\Pi}^{1/2}\bigr) = n - 1$.
Moreover,\break $\bolds{\Pi}^{1/2}
\mathbf{1}_n$ is eigenvector of both
$\bolds{\Pi}^{-1/2}\mathbf{L}\bolds{\Pi}^{-1/2}$ and $\bolds{\Pi}^{1/2}
\mathbf{H}_{\pi}' {\mathbf{L}^{+}}\mathbf{H}_{\pi} \bolds{\Pi}^{1/2}$ with associated
eigenvalue $0$. In
addition, if $\lambda\neq0$ is eigenvalue of $\bolds{\Pi}^{-1/2}
\mathbf{L}\bolds{\Pi}^{-1/2}$ with associated eigenvector
${\mathbf{u}}$, then
$\lambda^{-1}$ is eigenvalue of $\bolds{\Pi}^{1/2} \mathbf{H}_{\pi}'
{\mathbf{L}^{+}} \mathbf{H}_{\pi}\bolds{\Pi}^{1/2}$ with
associated eigenvector
$\mathbf{u}$. It thus follows from Lemma~\ref{LEM2} that~(\ref{EQRELAX6})
has the same solution as~(\ref{eqrelax2}) whenever $\mathbf{L}^{+} =
\mathbf{K}$.
As a result,~(\ref{eqrelax5}) has the same
solution as~(\ref{eqrelax1}).
\iffalse
\fi
\end{appendix}

\vspace*{-2pt}

\begin{thebibliography}{99}

%b1 ###
\bibitem[\protect\citeauthoryear{Bach and Jordan}{2006}]{Bach2006}
\textsc{Bach, F. R.} and \textsc{Jordan, M. I.} (2006).
Learning spectral clustering, with application to speech separation.
\textit{J. Mach. Learn. Res.} \textbf{7} 1963--2001.
\MR{2274430}

%b2 ###
\bibitem[\protect\citeauthoryear{Bartlett, Jordan and McAuliffe}{2006}]{BartlettJordanMcAuliffe2006}
\textsc{Bartlett, P. L., Jordan, M. I.} and \textsc{McAuliffe, J. D.} (2006).
Convexity, classification, and risk bounds.
\textit{J. Amer. Statist. Assoc.} \textbf{101} 138--156.
\MR{2268032}

%b3 ###
\bibitem[\protect\citeauthoryear{Belkin and Niyogi}{2002}]{Belkin2002}
\textsc{Belkin, M.} and \textsc{Niyogi, P.} (2002).
Laplacian eigenmaps and spectral techniques for embedding and clustering.
In \textit{Advances in Neural Information Processing Systems} \textbf
{14} 585--592. MIT Press, Cambridge, MA.

%b4 ###
\bibitem[\protect\citeauthoryear{Besag}{1974}]{Besag1974}
\textsc{Besag, J.} (1974).
Spatial interaction and statistical analysis of lattice systems (with
discussion).
\textit{J. Roy. Statist. Soc. Ser. B} \textbf{36} 192--236.
\MR{0373208}

%b5 ###
\bibitem[\protect\citeauthoryear{Besag and Kooperberg}{1995}]{BesagBio1995}
\textsc{Besag, J.} and \textsc{Kooperberg, C.} (1995).
On conditional and intrinsic autoregression.
\textit{Biometrika} \textbf{82} 733--746.
\MR{1380811}

%b6 ###
\bibitem[\protect\citeauthoryear{Carlin and Banerjee}{2003}]{CarlinBS2003}
\textsc{Carlin, B. P.} and \textsc{Banerjee, S.} (2003).
Hierarchical multivariate {CAR} models for spatio-temporally
correlated survival data (with discussion).
In \textit{Bayesian Statistics} \textbf{7} 45--63. Oxford Univ. Press.
\MR{2003166}

%b7 ###
\bibitem[\protect\citeauthoryear{Chan, Schlag and Zien}{1994}]{ChanCUT1994}
\textsc{Chan, P. K., Schlag, M. D. F.} and \textsc{Zien, J. Y.} (1994).
Spectral {K}-way ratio-cut partitioning and clustering.
\textit{IEEE Trans. Computer-Aided Design Integrated Circuits Syst.} \textbf{13} 1088--1096.

%b8 ###
\bibitem[\protect\citeauthoryear{Chung}{1997}]{Chung1997}
\textsc{Chung, F. R.} (1997).
\textit{Spectral Graph Theory}. Amer. Math. Soc., Providence, RI.
\MR{1421568}

%b9 ###
\bibitem[\protect\citeauthoryear{Dhillon, Guan and Kulis}{2007}]{DhillonPAMI2007}
\textsc{Dhillon, I. S., Guan, Y.} and \textsc{Kulis, B.} (2007).
Weighted graph cuts without eigenvectors: A multilevel approach.
\textit{IEEE Trans. Pattern Anal. Mach. Intell.} \textbf{29} 1944--1957.

%b10 ###
\bibitem[\protect\citeauthoryear{Ding, He and Simon}{2005}]{DingSDM2005}
\textsc{Ding, C., He, X.} and \textsc{Simon, H. D.} (2005).
On the equivalence of nonnegative matrix factorization and spectral
clustering.
In \textit{{SIAM} Conference on Data Mining (SDM)}. Newport Beach, CA.

%b11 ###
\bibitem[\protect\citeauthoryear{Donath and Hofmann}{1973}]{Donath1973}
\textsc{Donath, W. E.} and \textsc{Hofmann, A. J.} (1973).
Lower bounds for the partitioning of graphs.
\textit{IBM J. Res. Develop.} \textbf{17} 420--425.
\MR{0329965}

%b12 ###
\bibitem[\protect\citeauthoryear{Edelman, Arias and Smith}{1999}]{EdelmanSIAM1999}
\textsc{Edelman, A., Arias, T. A.} and \textsc{Smith, S. T.} (1999).
The geometry of algorithms with orthogonality constraints.
\textit{{SIAM} J. Matrix Anal. Appl.} \textbf{20} 303--353.
\MR{1646856}

%b13 ###
\bibitem[\protect\citeauthoryear{Fiedler}{1973}]{Fiedler1973}
\textsc{Fiedler, M.} (1973).
Algebraic connectivity of graphs.
\textit{Czechoslovak Math. J.} \textbf{23} 298--305.
\MR{0318007}

%b14 ###
\bibitem[\protect\citeauthoryear{Gower and Dijksterhuis}{2004}]{GowerBook2004}
\textsc{Gower, J. C.} and \textsc{Dijksterhuis, G. B.} (2004).
\textit{Procrustes Problems}. Oxford Univ. Press.
\MR{2051013}

%b15 ###
\bibitem[\protect\citeauthoryear{Guattery and Miller}{2000}]{GuatterySIAM2000}
\textsc{Guattery, S.} and \textsc{Miller, G. L.} (2000). Graph
embeddings and Laplacian eigenvalues.
\textit{{SIAM} J. Matrix Anal. Appl.} \textbf{21} 703--723.
\MR{1740868}

%b16 ###
\bibitem[\protect\citeauthoryear{Gupta and Nagar}{2000}]{GuptaNBook2000}
\textsc{Gupta, A. K.} and \textsc{Nagar, D. K.} (2000).
\textit{Matrix Variate Distributions}. Chapman and Hall, London.
\MR{1738933}

%b17 ###
\bibitem[\protect\citeauthoryear{Horn and Johnson}{1985}]{Horn1985}
\textsc{Horn, R. A.} and \textsc{Johnson, C. R.} (1985).
\textit{Matrix Analysis}. Cambridge Univ. Press.
\MR{0832183}

%b18 ###
\bibitem[\protect\citeauthoryear{Juh\'{a}sz and M\'{a}lyusz}{1977}]{Juhasz1977}
\textsc{Juh\'{a}sz, F.} and \textsc{M\'{a}lyusz, K.} (1977).
Problems of cluster analysis from the viewpoint of numerical analysis.
In \textit{Numerical Methods, Colloquia Mathematica
Societatis Janos Bolyai} (P. R\'{o}zsa, ed.) \textbf{22}
405--415. North-Holland,
Amsterdam.
\MR{0590916}

%b19 ###
\bibitem[\protect\citeauthoryear{Kannan, Vempala and Vetta}{2000}]{Kannan}
\textsc{Kannan, R., Vempala, S.} and \textsc{Vetta, A.} (2000).
On clusterings: Good, bad, and spectral.
In \textit{Proceedings of the 41st Annual Symposium on the Foundation of
Computer Science} 367--380.
\MR{1931834}

%b20 ###
\bibitem[\protect\citeauthoryear{Mardia}{1988}]{MardiaJMA1988}
\textsc{Mardia, K. V.} (1988).
Multi-dimensional multivariate Gaussian Markov random fields with
application to image processing.
\textit{J.~Multivariate Anal.} \textbf{24} 265--284.
\MR{0926357}

%b21 ###
\bibitem[\protect\citeauthoryear{Mardia, Kent and Bibby}{1979}]{Mardia1979}
\textsc{Mardia, K. V., Kent, J. T.} and \textsc{Bibby, J. M.} (1979).
\textit{Multivariate Analysis}. Academic Press, New York.
\MR{0560319}

%b22 ###
\bibitem[\protect\citeauthoryear{Meil\u{a} and Pentney}{2007}]{MeilaSAM2007}
\textsc{Meil\u{a}, M.} and \textsc{Pentney, W.} (2007).
Clustering by weighted cuts in directed graphs.
In \textit{{SIAM} Conference on Data Mining (SDM)}. Minneapolis, MN.

%b23 ###
\bibitem[\protect\citeauthoryear{Meil\u{a} and Shi}{2000}]{MeilaShi2000}
\textsc{Meil\u{a}, M.} and \textsc{Shi, J.} (2000).
Learning segmentation by random walks.
In \textit{Advances in Neural Information Processing} \textbf{12}
470--477. MIT Press, Cambridge, MA.

%b24 ###
\bibitem[\protect\citeauthoryear{Mohar}{1991}]{Mohar91}
\textsc{Mohar, B.} (1991). \textit{The Laplacian Spectrum of Graphs}
871--898. Wiley, New York.
\MR{1170831}

%b25 ###
\bibitem[\protect\citeauthoryear{Ng, Jordan and Weiss}{2002}]{Ng2002}
\textsc{Ng, A. Y., Jordan, M. I.} and \textsc{Weiss, Y.} (2002).
On spectral clustering: Analysis and an algorithm.
In \textit{Advances in Neural Information Processing Systems} \textbf{14} 849--856. MIT Press, Cambridge, MA.

%b26 ###
\bibitem[\protect\citeauthoryear{Rahimi and Recht}{2004}]{RahimiRecht2004}
\textsc{Rahimi, A.} and \textsc{Recht, B.} (2004).
Clustering with normalized cuts is clustering with a hyperplane.
In \textit{Workshop on Statistical Learning in Computer Vision}.
Prague, Czech Republic.

%b27 ###
\bibitem[\protect\citeauthoryear{Rand}{1971}]{RandRI1971}
\textsc{Rand, W. M.} (1971).
Objective criteria for the evaluation of clustering methods.
\textit{J. Amer. Statist. Assoc.} \textbf{66} 846--850.

%b28 ###
\bibitem[\protect\citeauthoryear{Ripley}{1981}]{Ripley1981}
\textsc{Ripley, B. D.} (1981). \textit{Spatial Statistics}. Wiley,
New York.
\MR{0624436}

%b29 ###
\bibitem[\protect\citeauthoryear{Saerens et al.}{2004}]{SaerensECML2004}
\textsc{Saerens, M., Fouss, F., Yen, L.} and \textsc{Dupont, P.} (2004).
The principal components analysis of a graph, and its relationships
to spectral clustering.
In \textit{The 15th European Conference on Machine Learning (ECML)} 371--383.

%b30 ###
\bibitem[\protect\citeauthoryear{Shen and Wang}{2007}]{ShenWang07}
\textsc{Shen, X.} and \textsc{Wang, L.} (2007).
Generalization error for multi-class margin classification.
\textit{Electron. J. Statist.} \textbf{1} 307--330.
\MR{2336036}

%b31 ###
\bibitem[\protect\citeauthoryear{Shi and Malik}{2000}]{Shi2000}
\textsc{Shi, J.} and \textsc{Malik, J.} (2000). Normalized cuts and
image segmentation.
\textit{{IEEE} Trans. Pattern Anal. Mach. Intell.} \textbf{22} 888--905.

%b32 ###
\bibitem[\protect\citeauthoryear{Shortreed and Meil\u{a}}{2005}]{MeilaUAI2005}
\textsc{Shortreed, S.} and \textsc{Meil\u{a}, M.} (2005).
Unsupervised spectral learning.
In \textit{Proceedings of the Twenty-First Conference on Uncertainty in
Artificial Intelligence} 534--541. AUAI Press.

%b33 ###
\bibitem[\protect\citeauthoryear{von Luxburg, Belkin and Bousquet}{2008}]{LuxburgTR2004}
\textsc{von Luxburg, U., Belkin, M.} and \textsc{Bousquet, O.} (2008).
Consistency of spectral clustering. \textit{Ann. Statist.} \textbf{36} 555--586.
\MR{2396807}

%b34 ###
\bibitem[\protect\citeauthoryear{von Luxburg}{2007}]{LuxburgSC2007}
\textsc{von Luxburg, U.} (2007).
A tutorial on spectral clustering.
\textit{Statist. Comput.} \textbf{17} 395--416.
\MR{2409803}

%b35 ###
\bibitem[\protect\citeauthoryear{Wahba}{1990}]{Wahba1990}
\textsc{Wahba, G.} (1990). \textit{Spline Models for Observational Data}.
SIAM, Philadelphia.
\MR{1045442}

%b36 ###
\bibitem[\protect\citeauthoryear{Webb}{2002}]{Webb2002}
\textsc{Webb, A. R.} (2002).
\textit{Statistical Pattern Recognition}, 2nd ed. Wiley, Hoboken.
\MR{2191640}

%b37 ###
\bibitem[\protect\citeauthoryear{Weiss}{1999}]{Weiss2001}
\textsc{Weiss, Y.} (1999).
Segmentation using eigenvectors: A unifying view.
In \textit{IEEE International Conference on Computer Vision} 975--982.
IEEE Computer Society.

%b38 ###
\bibitem[\protect\citeauthoryear{Yeung et al.}{2001}]{YeungBio2001}
\textsc{Yeung, K. Y., Fraley, C., Murua, A., Raftery, A. E.} and
\textsc{Ruzzo, W. L.} (2001).
Model-based clustering and data transformations for gene expression
data.
\textit{Bioinform.} \textbf{17} 977--987.

%b39 ###
\bibitem[\protect\citeauthoryear{Yu and Shi}{2003}]{YuICCV2003}
\textsc{Yu, S. X.} and \textsc{Shi, J.} (2003).
Multiclass spectral clustering.
In \textit{IEEE International Conference on Computer Vision} 313--319.
IEEE Computer Society.

%b40 ###
\bibitem[\protect\citeauthoryear{Zha et al.}{2002}]{ZhaNIPS2002}
\textsc{Zha, H., Ding, C., Gu, M., He, X.} and \textsc{Simon, H.} (2002).
Spectral relaxation for $k$-means clustering.
In \textit{Advances in Neural Information Processing Systems} \textbf
{14} 1057--1064. MIT Press, Cambridge, MA.\

%b41 ###
\bibitem[\protect\citeauthoryear{Zou, Zhu and Hastie}{2006}]{ZouZhuHastieTR2006}
\textsc{Zou, H., Zhu, J.} and \textsc{Hastie, T.} (2006).
The margin vector, admissible loss and multi-class margin-based
classifiers.
Technical report, Dept. Statistics, Stanford Univ.

\end{thebibliography}
\end{document}